\documentclass[
twocolumn,
superscriptaddress,
 amsmath,amssymb,
 aps,
 pra,
 floatfix,
]{revtex4-2}
\usepackage{xcolor}
\usepackage{stackengine}
\usepackage[normalem]{ulem}
\usepackage{graphicx}
\usepackage{array}
\usepackage{booktabs}
\usepackage{dcolumn}
\usepackage{bm}
\usepackage[colorlinks]{hyperref}
\usepackage[capitalize]{cleveref}
\hypersetup{%
	plainpages=true,
	breaklinks=true,
	hypertexnames=false,
	pageanchor=true,
	colorlinks=true,
	linkcolor={blue},
	citecolor={red},
	urlcolor={blue},
	anchorcolor={black}
}

\usepackage{physics}

\usepackage{soul}
\usepackage{accents}
\usepackage{float}
\usepackage{appendix}

\newcommand{\figref}[1]{\mbox{Fig.~\ref{#1}}}

\newcommand{\tabref}[1]{\mbox{Table~\ref{#1}}}
\newcommand{\secref}[1]{\mbox{Sec.~\ref{#1}}}

\newcommand{\appref}[1]{\mbox{App.~\ref{#1}}}
\renewcommand{\eqref}[1]{\mbox{Eq.~(\ref{#1})}}

\newcommand{\expec}[1]{\left\langle{#1}\right\rangle}
\newcommand{\expecsm}[1]{\langle{#1}\rangle}
\newcommand{\be}{\begin{equation}}
\newcommand{\ee}{\end{equation}}
\newcommand{\bea}{\begin{eqnarray}}
\newcommand{\eea}{\end{eqnarray}}

\definecolor{darkgreen}{rgb}{0.1, 0.3, 0.0}

\newcommand{\figpanel}[2]{Fig.~\hyperref[#1]{\ref*{#1}(#2)}}
\newcommand{\figpanels}[3]{Fig.~\hyperref[#1]{\ref*{#1}(#2)-(#3)}}
\newcommand{\figpanelNoPrefix}[2]{\hyperref[#1]{\ref*{#1}(#2)}}

\usepackage[]{changes}

\begin{document}

\title{Non-Hermitian Quantum Nonlinear Optics with Single Photons}

\author{Samuel Napoli}
\email{samunapoli67@gmail.com}
\affiliation{Dipartimento di Scienze Matematiche e Informatiche, Scienze Fisiche e  Scienze della Terra, Universit\`{a} di Messina, I-98166 Messina, Italy}

\author{Andrea Zappal\'a}
\affiliation{Dipartimento di Scienze Matematiche e Informatiche, Scienze Fisiche e  Scienze della Terra, Universit\`{a} di Messina, I-98166 Messina, Italy}

\author{Franco Nori}
\affiliation{Quantum Information Physics Theory Research Team, RIKEN Center for Quantum Computing, Wako-shi, Saitama, 351-0198, Japan}
\affiliation{Physics Department, The University of Michigan, Ann Arbor, Michigan 48109-1040, USA}

\author{Salvatore Savasta}
\affiliation{Dipartimento di Scienze Matematiche e Informatiche, Scienze Fisiche e  Scienze della Terra, Universit\`{a} di Messina, I-98166 Messina, Italy}

\author{Daniele Lamberto}
\affiliation{Dipartimento di Scienze Matematiche e Informatiche, Scienze Fisiche e  Scienze della Terra, Universit\`{a} di Messina, I-98166 Messina, Italy}
\affiliation{Quantum Information Physics Theory Research Team, RIKEN Center for Quantum Computing, Wako-shi, Saitama, 351-0198, Japan}

\date{\today}

\begin{abstract}

{Quantum nonlinear optics seeks to harness strong photon--photon interactions for scalable quantum technologies, although dissipative losses still pose a major barrier to near‑unity conversion efficiency. Here, we bridge non-Hermitian physics with the quantum nonlinear domain by exploiting perfect absorption to identify and optimize few-photon nonlinear processes. We theoretically investigate two circuit QED systems, operating in the light-matter ultrastrong coupling regime. The first (i) enables simultaneous two-atom excitations by single photons, while the second (ii) realizes the strong coupling between a single-photon and a two-photon Fock states.
We demonstrate that, since the strong optical nonlinearities cause quantum spectral features to emerge already at the level of linear response theory, the perfect absorption condition in $\abs{S_{11}}$ enables near-deterministic single-photon down-conversion into (i) a qubit--qubit-correlated pair and (ii) a two-photon pair. We show that the conversion efficiency can be systematically optimized through experimentally accessible parameters, both linked to the emergence of Hermitian subspaces within the effective non-Hermitian Hamiltonians. These findings position non-Hermitian engineering as a broadly applicable route to optimizing quantum devices at the single-photon level, even beyond circuit-QED platforms.}

\end{abstract}

\maketitle

\section{Introduction} 
\label{sec:Introduction}

Nonlinear optics, which encompasses processes such as frequency conversion \cite{Franken1961, AmstrongPR1962}, all-optical switching \cite{gibbs1985optical,Almeida2004Nature}, and optical modulation \cite{WangNature2018, MinzioniIOP2019, Sinatkas2021JAP}, has been a central area of optical science for several decades \cite{Shen1984, Agrawal2012, Boyd2020}. A particularly compelling frontier within this field is quantum nonlinear optics (QNLO), a regime 
in which the optical response of a system depends on the quantum state of the electromagnetic field \cite{Chang2014}.
 Among the different phenomena and processes within QNLO, spontaneous parametric conversion occupies a central role, as it is among the most widely exploited processes for generating non-classical states of light, including squeezed states and entangled photon pairs \cite{BurnhamPRL1970,WuPRL1986, walls2008quantum, ZhaoPRL2020, ZhangReviewSPD2021, NehraSci2022, QuesadaADvPPhot2022}. More broadly, QNLO is instrumental for a wide class of quantum technologies, notably single-photon transistors \cite{ChenScience2013,SunScience2018, TsiamisPRA2026}, deterministic all-optical quantum logic  \cite{Milburn1989,Kimble1998,HackerNature2016, StolzPRX2022}, and platforms for generating strongly correlated light-matter states \cite{ReinhardNaturePH2012, Chang2014, BlochNature2022}.

Superconducting quantum circuits have emerged as one of the most compelling platforms for realizing QNLO in the microwave domain \cite{YouNori2011, Devoret2013, GuSciRep2017,BlaisNaturePhys2020, BlaisRevModPhys2021, KrantzAppPhysRev2019, Oliver2020}. The strong Josephson nonlinearity intrinsic to these circuits underpins key phenomena such as photon blockade \cite{LangPRL2011}, single-photon Kerr effects \cite{RebicPRL2009, KirchmairNature2013}, spontaneous parametric down-conversion \cite{AbdoPRL2013, WangPRA2015, SvessonPRB2017, Chang2020PRX}, and nonclassical photon‑statistics signatures, as demonstrated by antibunching and engineered single‑photon emission \cite{HouckNature2007, BozyigitNatPhysc2011, PechalPRX2014, RollandPRL2019}. Furthermore, in these superconducting circuits, light-matter coupling strength can be pushed into the ultrastrong (USC) and deep-strong coupling (DSC) regimes even with a single or few quantum emitters \cite{FornPRL2010, YouNori2011, YoshiaraNaturePhys2017,ChenPhysRevA2017, FriskKockumNatureRev2019, FornRevModPhys2019, WeiQiinNoriPhysRep2024, TorrasColomaApplPhysLett2025}. In the USC regime, multi-excitation exchange processes \cite{GarzianoPRL2016, StassiPRA2017, KockumPRA2017} and few-photon parametric conversion \cite{CongPRA2020, Koshino2022PRR} have been theoretically predicted. A few of these effects have recently been observed experimentally \cite{ TomonagaNatureComm2025, WangNatureComm2025}.

However, despite these advances, a persistent obstacle across all QNLO platforms is the unavoidable presence of dissipation. Photon loss not only degrades the purity of quantum states but directly limits the efficiency of nonlinear conversion processes, which must approach unity for several applications. 
Analogously to classical nonlinear optics, where the concept of critical coupling (or impedance matching) is widely employed to optimize conversion efficiency \cite{Cai2000PRL, Guo2016PRL, BruchAppPhysLett2018,ZhiYan2025}, early proposals in the quantum domain demonstrated single-photon down-conversion by exploiting the same principle in circuit and waveguide QED architectures, typically involving three-level emitters in $\Lambda$ or $\Delta$ configurations \cite{Koshino2009PRA, Inomata2014PRL, SanchezPRA2016}.

Non-Hermitian physics offers a powerful and complementary perspective on this challenge. It has been established that tailored gain-loss distributions (or, remarkably, purely lossy configurations) can be exploited to reshape wave propagation, giving rise to a rich phenomenology in classical photonic systems. In particular, in platforms with parity-time (PT) symmetry \cite{El-GanainyNaturePhys2018, AshidaAdvPhy2020, WangReviewOptica2023}, striking effects such as loss-induced transparency, unidirectional invisibility, and enhanced sensing near exceptional points have been demonstrated  \cite{GuoPRL2009, LinPRL2011, ChenNature2017, WiersigOPTICA2020}. 
Beyond the PT-symmetric case, non-Hermiticity also governs wave-scattering anomalies such as coherent perfect absorption (CPA) \cite{WanScience2011,Zanotto2014Nature, zhangNATCOMM2017} and perfect absorption (PA), often referred to as reflectionless scattering modes \cite{SweeneyPhysRevA.102.063511,ImaniAdFuncMat2020, Rao2024NaturePhysics, JiangNature2024}, in which all incoming radiation is absorbed without reflection. In open passive systems, PA is directly connected to the eigenspectrum of an effective non-Hermitian Hamiltonian and represents a spectral impedance-matching condition that can be precisely engineered \cite{SweeneyPhysRevA.102.063511,HanScienceAdv2023, Rao2024NaturePhysics,JiangNature2024, BonizzoniNatureComm2025}. However, transposing these ideas to the QNLO domain (where few-photon states, quantum correlations, and non-classical statistics are the central objects of interest) remains largely unexplored.

In this work, we bridge concepts from non-Hermitian photonics with the QNLO domain. In particular, we theoretically investigate PA (and its connection with non-Hermitian physics) within a fully quantum framework under a weak continuous-wave coherent drive, and study its impact on the maximal achievable conversion efficiency. As realistic examples, we consider two circuit QED setups: (i) two quantum emitters (either identical or non-identical) coupled to a microwave resonator in the USC regime, and (ii) a system in which single- and two-photon states are strongly coupled through LC resonators connected to a common flux qubit acting as a nonlinear coupler. The spectral features associated with the underlying quantum nonlinear processes for both setups have recently been observed experimentally \cite{ TomonagaNatureComm2025, WangNatureComm2025}. For setup (ii), however, we adopt an alternative configuration to allow for the independent tuning of the relevant parameters. In the few-photon regime, the strong nonlinearities of these systems cause quantum spectral features to emerge already at the level of linear response theory, so that the underlying nonlinear processes are directly encoded in the linear reflection spectrum $\abs{S_{11}}$. 
Although absorption is traditionally viewed as a detrimental mechanism leading to signal degradation, here we exploit PA to spectrally identify the conditions leading to highly efficient quantum nonlinear effects. Specifically, in the two-qubit-resonator setting, we demonstrate that this effect enables near-deterministic single-photon down-conversion from the resonator input channel to the qubit output channels, with the emitted photons forming a qubit-qubit-correlated pair. 
While strictly unity efficiency requires the absence of non-radiative losses, we further show that even in the presence of realistic dissipation the efficiency can be systematically optimized by tuning experimentally accessible circuit parameters. This behavior can be linked to the emergence of Hermitian subspaces within the effective non-Hermitian Hamiltonian governing the spectral features \cite{BonizzoniNatureComm2025}. 
Furthermore, we analyze the correlation properties of the output fields and their non-classical signatures, and extend this study to the weak nonlinear coupling regime. We also provide a detailed analysis of the impact of the relevant parameters (nonlinear coupling strengths, radiative and non radiative losses, detuning) on the conversion efficiency.

This paper is organized as follows. In \secref{onephtwatoms}, we introduce the effective Hamiltonian describing the single-photon--two-atom excitation process. In \secref{theory_s}, we present the scattering matrix formalism used to compute the reflection spectrum $\abs{{S_{11}}}$, and its connection to PA and the non-Hermitian Hamiltonian $H_{\rm PA}$. Specifically, in \secref{resPT}, we analyze the system (i) under the PT-symmetry condition, evaluating the role of non-radiative losses on conversion efficiency and the correlation properties of the emitted fields. In \secref{Hrzsubsp}, we show how light-matter detuning, in the absence of loss balance, induces a Hermitian subspace in $H_{\mathrm{PA}}$ and discuss its impact on efficiency and non-classical signatures, demonstrating the robustness and applicability of our results even beyond PT symmetry. In \secref{weakeffcoupl}, we extend the analysis to the weak nonlinear coupling regime. Section\,\ref{phasediagrams} maps the maximum achievable conversion efficiency across a wide parameter space using phase diagrams. Finally, in \secref{SCfockstates}, we apply our framework to a purely photonic setting (ii) to investigate near-deterministic single-to-two-photon down-conversion.

\section{One Photon Exciting Two Atoms}\label{onephtwatoms}
\begin{figure}[t]
\includegraphics[width= \linewidth]{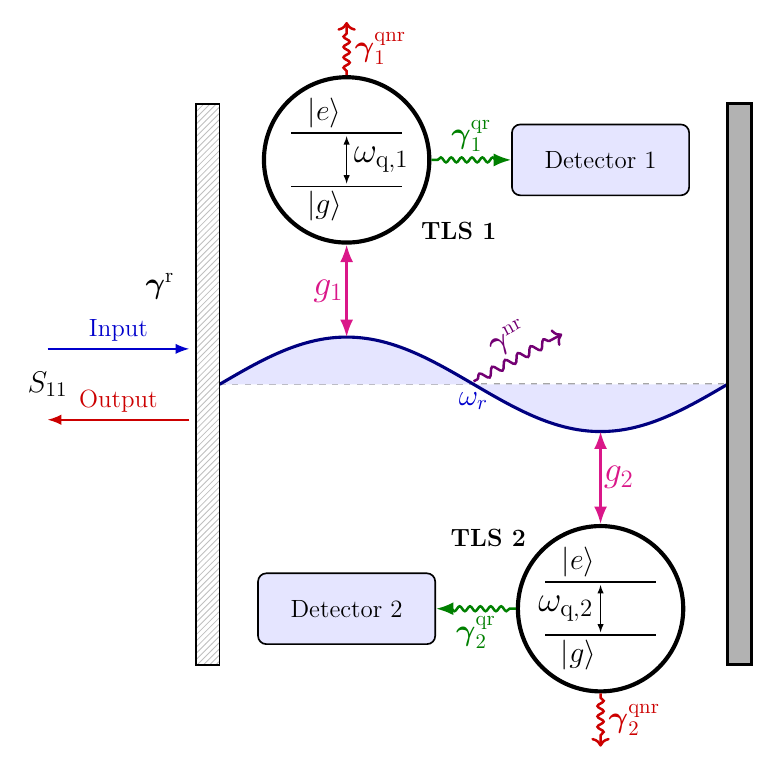}
\caption{Sketch of the circuit-QED setup hosting the quantum nonlinear process in which two qubits can be simultaneously excited by a single photon, as described by the Hamiltonian in \eqref{Hams}. 
For illustrative purposes, the resonator is represented as a waveguide cavity coupled to two individual qubits, depicted as circles enclosing two energy levels, and placed at opposite antinodes of the cavity mode. Radiative decay arrows for the resonator and qubits ($\gamma^{\mathrm{r}}$ and $\gamma^{\mathrm{qr}}_{1,2}$) point towards the input mirror and respective qubit detectors, while intrinsic non-radiative loss arrows ($\gamma^{\mathrm{nr}}$ and $\gamma^{\mathrm{qnr}}_{1,2}$) point outwards to the environment. The radiative decay channel of the resonator serves simultaneously as the input drive port and the output collection channel.}

\label{fig_modello_fin}
\end{figure}

We start by considering the circuit QED setup composed of two flux qubits coupled to a common LC resonator, as sketched in \cref{fig_modello_fin}. This setup allows for the observation of the nonlinear process in which a single photon is able to simultaneously excite two atoms, as recently realized in Ref.\,\cite{TomonagaNatureComm2025}. Setting $\hbar = 1$, this system can be modeled by the generalized Dicke Hamiltonian:
\begin{equation}\label{Hams}
\begin{split}
{H}_S = \sum_{j=1}^2 & \frac{\omega_{{\rm q},j}}{2}\,\sigma^{z}_j + \omega_{r}\, a^\dagger a 
- (g_1\Lambda_1 - g_2\Lambda_2)\, (a + a^\dagger) \\
& - 2\, \frac{g_1 g_2}{\omega_{r}}\, \Lambda_1 \Lambda_2 \,,
\end{split}
\end{equation}
where $\omega_{r}$ is the resonator frequency, $a$ ($a^\dagger$) are the corresponding bosonic annihilation (creation) operators, $\sigma_j^\alpha$ are Pauli matrices of the $j$-th qubit (with $\alpha\in\{x,y,z\}$), and $\Lambda_j = \cos{\theta_j}\, \sigma^x_j + \sin{\theta_j}\, \sigma^z_j$ denotes the qubits' operator responsible for the coupling with the resonator (expressed in the qubit energy basis). 
The mixing angle is given by $\theta_j = -\arctan(\epsilon_j / \Delta_j)$, where $\Delta_j$ is the qubit gap at zero bias, $\epsilon_j$ is the flux offset, and $\omega_{{\rm q},j} = \sqrt{\epsilon_j^2 + \Delta_j^2}$. 
This configuration introduces a longitudinal interaction term whenever $\theta_j \neq 0$, which induces the system parity-symmetry breaking.
Nonetheless, for $\theta_j =0$, we observe that the Hamiltonian in \eqref{Hams} closely resembles the cavity-QED dipole-gauge Hamiltonian, describing the interaction between two natural atoms and a cavity mode \cite{DeBernardisPRA2018,GarzianoPRA2020, lamberto2024superradiantquantumphasetransition, ZappalaPRA2025}. The only difference between these two lies in the opposite sign of the coupling strengths $g_1$ and $g_2$, which naturally arises from the circuit geometry, as the two qubits have a different phase in their coupling with the shared resonator. 
For such a reason, in the sketch in \figref{fig_modello_fin}, the qubits are placed in two different antinodes with opposite phases. 
This platform enables experimental access to the USC regime with two quantum emitters, while allowing for a controllable symmetry-breaking mechanism through the variation of the flux offsets $\epsilon_j$.

\begin{figure}[t]
\includegraphics[width=\linewidth]{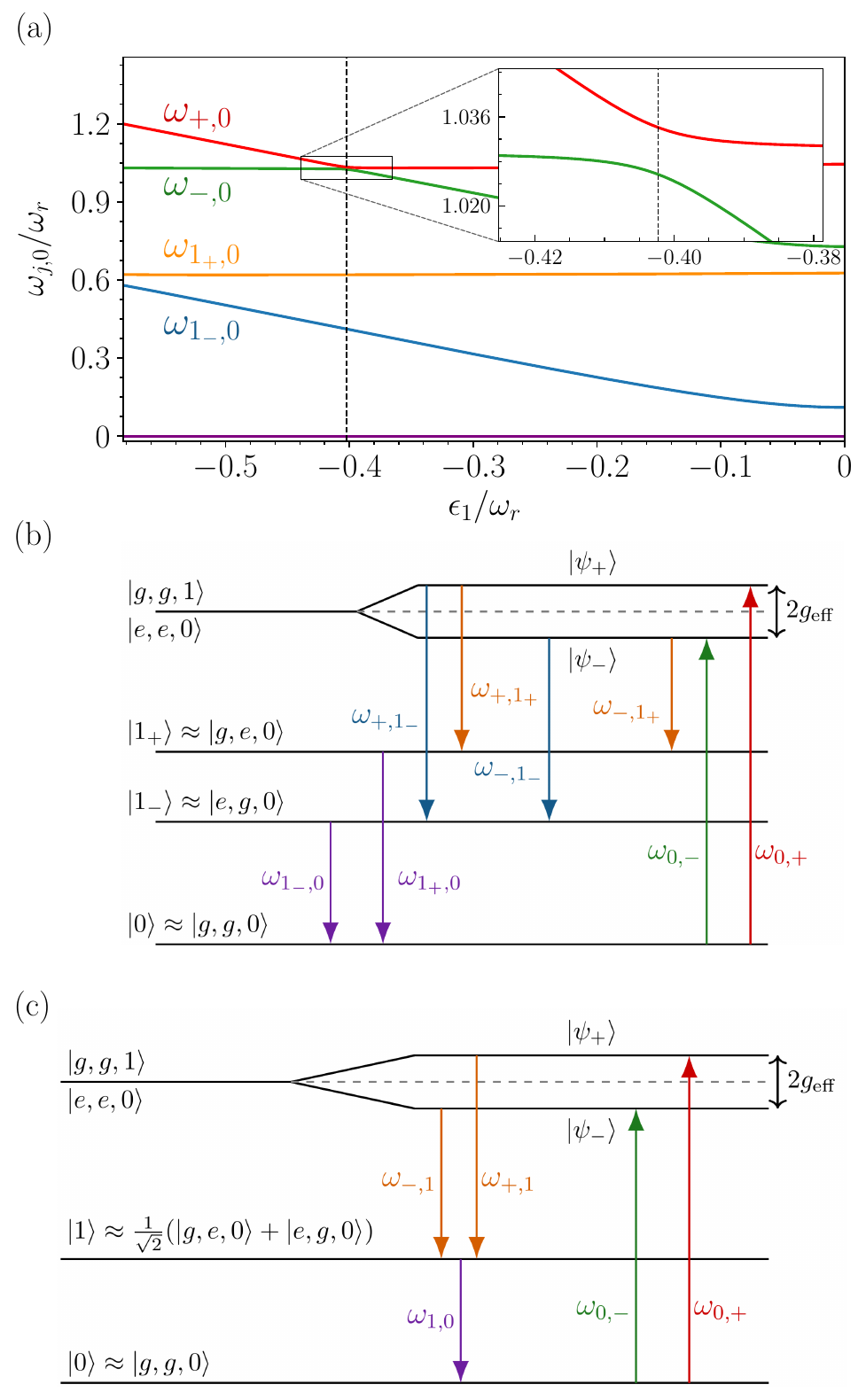}
\caption{Energy spectrum and schematic transitions diagrams of the single-photon two-qubit excitation process. (a) Eigenfrequencies of $H_{S}$ as a function of flux offset $\epsilon_1$, both normalized by the resonator frequency $\omega_{r}$. Here, $\omega_{j,0}$ denotes the transition frequency with respect to the ground state. The inset provides a magnified view of the anticrossing region near $\omega_{\mathrm{r}} \approx \omega_{\mathrm{q},1} + \omega_{\mathrm{q},2}$, which originates from the effective coupling between the $|g,g,1\rangle$ and $|e,e,0\rangle$ states. The parameters used for this panel are the same as those in Ref.\,\cite{TomonagaNatureComm2025}. (b) Energy levels for non-identical qubits at zero detuning, $\Delta_{\omega} = \tilde{\omega}_{\mathrm{r}} - (\tilde{\omega}_{\mathrm{q},1} + \tilde{\omega}_{\mathrm{q},2}) = 0$, marked by the vertical dashed line in panel (a). (c) Energy level scheme for identical qubits ($\tilde{\omega}_{\mathrm{q},1} = \tilde{\omega}_{\mathrm{q},2} \equiv \tilde{\omega}_{\mathrm{q}}$). 
Upward arrows indicate the coherent excitation from the ground state $|0\rangle$ to the hybrid modes $|\psi_{\pm}\rangle$, while downward arrows illustrate spontaneous emission processes induced by the qubits' losses. $\omega_{j,k}$ denotes the transition frequency from initial state $j$ to final state $k$.
}
\label{fig_avoided_crossing}
\end{figure}

In the USC regime and for $\theta_j \neq 0$, the energy spectrum of the Hamiltonian in \eqref{Hams} displays a characteristic avoided level crossing at $\omega_r \approx \omega_{\text{q},1} + \omega_{\text{q},2}$, as shown in \figref{fig_avoided_crossing}(a) and its inset. This spectral feature emerges from an effective coupling between the states $|g,g,1\rangle$ and $|e,e,0\rangle$ mediated by energy-nonconserving virtual transitions, driven by the counter-rotating terms in \eqref{Hams}. Notably, this effect occurs even with equal signs of the coupling strengths, and in the absence of spin-spin interaction, namely the last term in \eqref{Hams} \cite{KockumPRA2017,GarzianoPRL2016}. The time-resolved dynamics of the ``one-photon-exciting-two-atoms'' phenomenon has been theoretically investigated in Ref.~\cite{GarzianoPRL2016}, where vacuum Rabi oscillations between one photon in the resonator and the two qubits' excitations were predicted.

It has been shown that this quantum nonlinear process can be described by the following effective Hamiltonian \cite{GarzianoPRL2016, KockumPRA2017}:
\begin{equation}\label{Heff}
H_{\rm eff} = \sum_{j=1}^2  \frac{\tilde\omega_{{\rm q},j}}{2}\,\sigma^{z}_j + \tilde\omega_{r}\, a^\dagger a +  g_{\rm eff}(a \,\sigma^+_1 \sigma^+_2+\rm H.c.)\,,
\end{equation}
where $\sigma^{+}_j = (\ket{e}\bra{g})_j$ and $\sigma^{-}_j = (\sigma^{+}_j)^\dagger$ are the transition operators for the $j$-th qubit. The effective resonant frequencies of the qubits and resonator are denoted as $\tilde\omega_{{\rm q},j}$ and $\tilde\omega_{r}$, respectively, each dressed by the USC and spin-spin interaction in \eqref{Hams}.

In \appref{g_eff_app}, we employ perturbation theory to calculate $g_{\rm eff}$, explicitly accounting for the spin-spin interaction. This term introduces additional virtual transitions which contribute to the effective coupling, not included in previous works \cite{GarzianoPRL2016,TomonagaNatureComm2025}. In particular, these additional pathways significantly reduce the magnitude of $g_{\rm eff}$ compared to the case in which this interaction is absent. However, in both scenarios, $g_{\rm eff}$ still remains much smaller than the dressed frequencies ($g_{\rm eff} \ll \tilde\omega_{r}, \,\tilde\omega_{{\rm q},j}$). Hence, in the weak excitation regime, we can safely restrict $H_{\rm eff}$ to the coherent exchange between a single photon and two atomic excitations.

Since $H_{\text{eff}}$ commutes with the generalized excitation number operator $N_{\text{exc}} = 2a^\dagger a + \sigma_1^+\sigma_1^- + \sigma_2^+\sigma_2^-$, the Hilbert space naturally block-diagonalizes into independent excitation manifolds with constant $N_{\text{exc}}$. Thus, the ground state $|0\rangle \approx |g,g,0\rangle$ spans the $N_{\text{exc}}=0$ manifold. If the qubits feature distinct transition frequencies, in the vicinity of the avoided crossing, the $N_{\text{exc}}=1$ sector is well approximated by the states $\ket{1_-}\approx|e,g,0\rangle$ and $\ket{1_+}\approx|g,e,0\rangle$, as illustrated by the energy levels in \figref{fig_avoided_crossing}(b). 
To simplify the discussion, however, we will primarily focus on identical qubits ($\tilde{\omega}_{\text{q},1} = \tilde{\omega}_{\text{q},2} \equiv \tilde{\omega}_{\text{q}}$) throughout the main text. 
Under this assumption, the $N_{\text{exc}}=1$ subspace can be conveniently described by the 
symmetric $|1\rangle \approx (|e,g,0\rangle 
+ |g,e,0\rangle)/{\sqrt{2}}$ and antisymmetric $(|e,g,0\rangle - |g,e,0\rangle)/{\sqrt{2}}$ eigenstates, as shown in \figref{fig_avoided_crossing}(c). However, the latter does not contribute the spectral features investigated in the following sections.
The impact of non-identical qubits on the system spectral features will be addressed in \appref{appD}.

The two-atom excitation process previously described occurs entirely within the $N_{\rm exc}=2$ manifold, which is spanned by the states $\{\ket{g,g,1}, \ket{e,e,0}\}$. Diagonalizing $H_{\rm eff}$ in this two-state subspace gives the upper $(\omega_+)$ and lower $(\omega_-)$ hybrid-mode eigenfrequencies
\begin{equation}\label{eigenpclos}
\omega_{\pm} = \frac{1}{2}\biggl(\tilde\omega_{r} \pm \sqrt{4g^2_{\rm eff}+\Delta^2_\omega}\biggr)\,,
\end{equation}
where $\Delta_\omega= \tilde \omega_{r}-2 \tilde\omega_{\rm q}$ is the total resonator-qubit detuning. The corresponding hybrid-mode eigenstates $\ket{\psi_\pm}$ are given by
\begin{align}
\label{eq4}
\ket{\psi_+}= & \quad\cos{\biggl(\frac{\alpha}{2}\biggr)} \ket{g,g,1} + \sin{\biggl(\frac{\alpha}{2}\biggr)} \ket{e,e,0}\,,\\
\ket{\psi_-} =&- \sin{\biggl(\frac{\alpha}{2}\biggr)} \ket{g,g,1}+ \cos{\biggl(\frac{\alpha}{2}\biggr)} \ket{e,e,0} \,, \label{eq5}
\end{align}
with $\tan{(\alpha)} = 2g_{\rm eff}/\Delta_\omega$ and $\alpha \in [0, \pi]$.

\section{Spectra, emission properties, and correlations}\label{theory_s}

In this section, we first present the expression for the reflection spectrum in the weak-excitation regime for the system described in the previous section, with particular emphasis on the connection between PA and the underlying non-Hermitian physics. We begin in \cref{resPT} by discussing the case of a PT-symmetric system in the strong-coupling (SC) regime, showing how the proximity to PA can significantly enhance the qubits' emission efficiencies, as well as the non-classical properties of the emitted excitations. Subsequently, \cref{Hrzsubsp} relaxes the PT-symmetry requirement, demonstrating that the presence of Hermitian subspaces (which can be obtained even in the absence of loss balance) is sufficient to enhance the previously predicted features. Finally, in \cref{weakeffcoupl}, we investigate the presence of PA and its impact on the emission properties when the system operates in the weak coupling regime.

Considering the system described in \cref{onephtwatoms}, we assume a weak coherent microwave field driving the resonator. The reflection spectrum is extracted from the reflected signal by coupling the resonator to an open semi-infinite transmission line (TL). To detect the output fields from the two qubits, each of them is additionally connected to a semi-infinite TL, as shown schematically in \figref{fig_modello_fin}. The parameters $\gamma^{\rm r}$ and $\gamma^{\rm qr}_{i}$ denote the radiative decay rates of the resonator and $i$-th qubit, respectively, arising from their coupling to the input–output TLs, while $\gamma^{\rm nr}$ and $\gamma^{\rm qnr}_{i}$ represent the corresponding non-radiative losses, arising from additional interactions with the external environment. These loss rates refer to processes within the specific energy subspaces illustrated in \figref{fig_avoided_crossing}(b,\,c). 
Hereafter, we focus on the $N_\mathrm{exc}=2$ manifold, since it is the sector in which the coherent exchange between the photonic and qubit excitations takes place.

Using standard input--output theory \cite{gardiner2004quantum,walls2008quantum} (see \appref{appA} for details), the positive-frequency component of the output field for each channel is related to the corresponding input field. Namely, we obtain the following input--output equations for the resonator and the qubits, respectively:
\begin{align}
    a^{(\rm r)}_{\rm out}(\omega_d) &= a^{(\rm r)}_{\rm  in}(\omega_d) - \sqrt{ \gamma^{\rm r}}\, a(\omega_d) \, ,\label{inoutres} \\
    b^{ \,(\rm r)}_{j,{\rm out}}(\omega_d) &= b^{\,(\rm r)}_{j,{\rm in}}(\omega_d) + \sqrt{\gamma^{\rm qr}_{j}}\, \sigma^{-}_{j}(\omega_d)\,, \label{inoutqs}
\end{align}
where $\omega_d$ is the drive frequency.
Since only the resonator is coherently driven, we have $|\langle a^{\rm (r)}_{\rm in} \rangle|^2 = |A_d|^2$, where $|A_d|^2$ represents the input photon rate, while $\langle b^{(\rm r)}_{j,{\rm in}} \rangle =0$. By employing the quantum-Langevin-equation (QLE) approach in the weak-excitation regime, in \appref{appA} we derive an analytical formula for the reflection spectrum at zero temperature:
\begin{equation}\label{s11}
\abs{S_{11}(\omega_d)} = \abs{\frac{\expec{a^{(\rm r)}_{\rm out}(\omega_d)}}{\expec{a^{( \rm r)}_{\rm in}(\omega_d)}}}
= \left| \frac{(\omega_d-\tilde \Omega_+)(\omega_d-\tilde \Omega_-)}{(\omega_d-\Omega_+)(\omega_d-\Omega_-)} \right| \, ,
\end{equation}
where $\tilde{\Omega}_j$ and $\Omega_j$ denote the zeros and poles of $\abs{S_{11}}$, respectively.
Explicit expressions for both these quantities are given in \appref{appA}.
In the SC regime, the real parts of these complex frequencies correspond to the transitions between the closed-system eigenfrequencies, i.e., the hybrid-mode eigenfrequencies derived in \eqref{eigenpclos} and the ground-state energy $-\tilde \omega_{\rm q}$.
The incoherent thermal contributions, neglected here, can be experimentally filtered out \cite{WangNatComm2023,WangNatureComm2025} by employing a coherent detection scheme (e.g., lock-in or homodyne), which selects only the signal component phase-locked to the drive. As a result, for this type of measurements, a millikelvin bath temperature would not significantly affect the coherent reflection (or transmission) spectra presented.

Following the approach commonly adopted for collective excitations coupled to electromagnetic resonators \cite{SweeneyPhysRevA.102.063511,HanScienceAdv2023,Rao2024NaturePhysics,BonizzoniNatureComm2025}, the poles $\Omega_j$ can be formally identified as the complex eigenvalues of the following effective non-Hermitian Hamiltonian
\begin{equation}\label{hres}
H_{\rm pol} = A -\frac{i}{2}\Gamma_{\rm pol}\,.
\end{equation}
Here, $A$ is the Hamiltonian corresponding to the closed-system transitions between the $N_\mathrm{exc} = 2$ manifold of $H_\mathrm{eff}$ and its ground state, while $\Gamma_{\rm pol}$ is dissipation matrix, given by, respectively:
\begin{align}
A &= \begin{pmatrix} \tilde \omega_{r} & g_{\rm eff} \\ g_{\rm eff} & 2\,\tilde\omega_{{\rm q}}  \end{pmatrix} \,,\label{A_matrix}\\
\Gamma_{\rm pol} &= \begin{pmatrix}  \gamma^{\rm r} +  \gamma^{\rm nr} & 0 \\ 0 &  \gamma^{\rm qr} + \gamma^{\rm qnr} \end{pmatrix}\,, \label{dissmatrix}
\end{align}
where $\gamma^{\rm qr\,(\rm qnr)}= \gamma^{\rm qr\,(\rm qnr)}_1+ \gamma^{\rm qr\,(\rm qnr)}_2$ represents the total radiative (non-radiative) qubits' loss. On the other hand, the complex frequencies $\tilde{\Omega}_{\pm}$ encode the conditions for PA, and are the eigenvalues of the non-Hermitian Hamiltonian 
\begin{equation}\label{hrz_bare}
H_{\mathrm{PA}} = A - \frac{i}{2}{\Gamma_{\rm PA}}\,,
\end{equation}
where $\Gamma_{\rm PA}$ is obtained from $\Gamma_{\rm pol}$ by inverting the sign of the input-output channel, namely, replacing $ \gamma^{\rm r}$ with $- \gamma^{\rm r}$. 
Specifically, when one of the eigenvalues $\tilde\Omega_{\pm}$ of $H_{\mathrm{PA}}$ becomes purely real (i.e., $\Im(\tilde\Omega_j)=0$ for $j$ either the upper or lower hybrid mode), it is then possible to tune the drive frequency to $\omega_d = \tilde{\Omega}_j$, such that $\abs{S_{11}}=0$ and thus realizing PA \cite{{SweeneyPhysRevA.102.063511,HanScienceAdv2023,Rao2024NaturePhysics,JiangNature2024,BonizzoniNatureComm2025}}.
Due to its sign-reversing, the resonator's radiative loss $\gamma^{\rm r}$ can be regarded as an effective gain in the dynamics of the zeros \cite{zhangNATCOMM2017}.

To better understand the physical meaning of the eigenfrequencies of $H_\mathrm{PA}$, we assume the SC regime (when $g_{\rm eff}$ is significantly larger than the loss rates, but still smaller than the dressed frequencies), and subsequently move to the hybrid-mode basis (see \appref{appA} for details). In this representation, the complex frequencies $\tilde\Omega_{\pm}$ correspond to the diagonal elements of the effective Hamiltonian
\begin{equation}\label{Hrz}
H_{\mathrm{PA}} = 
\begin{pmatrix}
\bar\Omega_1-\frac{i}{2}\tilde\gamma_1& 0 \\
0 &\bar\Omega_2-\frac{i}{2}{{\tilde\gamma_2}}
\end{pmatrix}\,,
\end{equation}
where the frequencies $\bar \Omega_{j}$ are the eigenvalues of the matrix $A$. The effective loss terms $\tilde{\gamma}_j$ read
\begin{equation}\label{gammabar}
\tilde{\gamma}_j = (- \gamma^{\rm r} +  \gamma^{\rm nr})|C_{j1}|^2 + (\gamma^{\rm qr} + \gamma^{\rm qnr})|C_{j2}|^2\,,
\end{equation}
where $j=1,2$ label the upper and lower hybrid modes, respectively. The Hopfield coefficients $|C_{jk}|^2$ (see Table~\ref{table1}) encode the photonic ($k=1$) and matter ($k=2$) fractions of each hybrid mode and depend on the detuning $\Delta_\omega$ through the mixing angle $\alpha$, defined in \eqref{eq4}.

A Hermitian subspace of $H_{\mathrm{PA}}$ emerges when an eigenvalue $\tilde{\Omega}_j$ (for some $j$) becomes purely real, i.e., when $\tilde{\gamma}_j = 0$. By tuning $\Delta_\omega$, one can change the weights of the Hopfield coefficients and shift the balance between the photonic and matter components of the hybrid modes, offering an efficient pathway to drive $\tilde{\gamma}_j$ to zero \cite{BonizzoniNatureComm2025}. As it is discussed in greater details in \secref{resPT}, the special case of PT symmetry in $H_{\mathrm{PA}}$ is realized when $\tilde{\omega}_r = 2\tilde{\omega}_{\rm q}$ and $\gamma^{\rm r} =  \gamma^{\rm nr} +  \gamma^{\rm qr} + \gamma^{\rm qnr}$. Under these conditions, $|C_{jk}|^2 = 1/2$ for all $j,k$, so that $\tilde{\gamma}_1 = \tilde{\gamma}_2 = 0$ and $H_{\mathrm{PA}}$ has an entirely real spectrum in its PT-unbroken region, thus enabling the presence of PA for both hybrid modes simultaneously \cite{zhangNATCOMM2017}.

\begin{table}[t]
\setlength{\lightrulewidth}{0.03em}
\centering
\fontsize{8}{9.6}\selectfont
\begin{tabular}{@{}c@{\hspace{0.35cm}}c@{}}
    \toprule
    \toprule
    Physical quantity & Symbol / Definition 
    \\
    \addlinespace[0.7mm]
    \toprule
    \addlinespace[1.0mm]

    \begin{tabular}[c]{@{}c@{}}
    Bare $j$-th qubit (resonator)\\resonance frequency
    \end{tabular} &
    $\omega_{{\rm q}, j} \,\, \left( \omega_r \right)$
    \\
    \addlinespace[0.2mm]
    \midrule
    \addlinespace[1.0mm]

    \begin{tabular}[c]{@{}c@{}}
    Effective $j$-th qubit (resonator)\\resonance frequency
    \end{tabular} &
    $\tilde{\omega}_{{\rm q}, j} \,\, \left( \tilde{\omega}_r \right)$
    \\
    \addlinespace[0.2mm]
    \midrule
    \addlinespace[1.0mm]

    Resonator--qubits detuning &
    $\Delta_\omega = \tilde \omega_{r} - \sum_j \tilde\omega_{{\rm q}, j}$
    \\
    \addlinespace[0.2mm]
    \midrule
    \addlinespace[1.0mm]

    \begin{tabular}[c]{@{}c@{}}
    Eigenfrequencies of $H_\mathrm{eff}$\\in $N_\mathrm{exc} \!=\! 2$ manifold 
    \end{tabular} &
    $\omega_{\pm} = \frac{1}{2}\!\left(\tilde\omega_{r} \pm\! \sqrt{4g^2_{\rm eff}+\Delta^2_\omega}\right)$
    \\
    \addlinespace[0.2mm]
    \midrule
    \addlinespace[1.0mm]

    Hopfield coefficients &
    $|C_{jk}|^2 \!=\! \begin{cases}
        \cos^2(\alpha/2) \,\,\text{for} \,\,j \!=\! k \\
        \sin^2(\alpha/2) \,\,\,\text{for} \,\,j \!\neq\! k
    \end{cases}$
    \\
    \addlinespace[0.2mm]
    \midrule
    \addlinespace[1.0mm]
    
    \begin{tabular}[c]{@{}c@{}}
    $j$-th qubit radiative\\(non-radiative) losses
    \end{tabular} &
    $\gamma_j^\mathrm{qr\,(qnr)}$
    \\
    \addlinespace[0.2mm]
    \midrule
    \addlinespace[1.0mm]
    
    \begin{tabular}[c]{@{}c@{}}
    Total qubits radiative\\(non-radiative) losses
    \end{tabular} &
    $\gamma^\mathrm{qr\,(qnr)} = \sum_j \gamma_j^\mathrm{qr\,(qnr)}$
    \\
    \addlinespace[0.2mm]
    \midrule
    \addlinespace[1.0mm]
    
    \begin{tabular}[c]{@{}c@{}}
    Resonator radiative\\(non-radiative) losses
    \end{tabular} &
    $\gamma_\mathrm{r\,(nr)}$
    \\
    \addlinespace[0.2mm]
    \midrule
    \addlinespace[1.0mm]

    \begin{tabular}[c]{@{}c@{}}
    Effective non-Hermitian\\Hamiltonian for PA (poles)
    \end{tabular} &
    $H_{\mathrm{PA(pol)}} = A - \frac{i}{2}{\Gamma_{\rm PA (pol)}}$
    \\
    \addlinespace[0.2mm]
    \midrule
    \addlinespace[1.0mm]

    Eigenvalues of $H_{\mathrm{PA}}$ in SC &
    $\tilde{\Omega}_j \approx \bar\Omega_j-\frac{i}{2}\tilde\gamma_j$
    \\
    \addlinespace[0.2mm]
    \midrule
    \addlinespace[1.0mm]
    
    Effective hybrid-mode losses &
    $\begin{aligned}
        \tilde{\gamma}_j =&\, (- \gamma^{\rm r} +  \gamma^{\rm nr})|C_{j1}|^2 \\
        &+(\gamma^{\rm qr} + \gamma^{\rm qnr})|C_{j2}|^2
    \end{aligned}$
    \\
    \addlinespace[0.2mm]
    \midrule
    \addlinespace[1.0mm]
    
    Reflection spectrum &
    $|S_{11}(\omega)|$
    \\
    \addlinespace[0.2mm]
    \midrule
    \addlinespace[1.0mm]
    
    $j$-th qubit emission spectrum &
    $\tilde{S}_{{\rm q},j}(\omega)$
    \\
    \addlinespace[0.2mm]
    \midrule
    \addlinespace[1.0mm]

    \begin{tabular}[c]{@{}c@{}}
    Qubit--qubit\\cross-correlation spectrum
    \end{tabular} &
    ${\cal S}_{j,k}(\omega)$
    \\
    \addlinespace[0.2mm]
    \midrule
    \addlinespace[1.0mm]

    $j$-th qubit emission efficiency &
    $\eta_j (\omega)$
    \\
    \addlinespace[0.2mm]
    \midrule
    \addlinespace[1.0mm]

    \begin{tabular}[c]{@{}c@{}}
    Normalized two-qubit\\output correlation function
    \end{tabular} &
    $C^q_2(\omega)$
    \\
    \addlinespace[0.2mm]
    
    \bottomrule
    \bottomrule
\end{tabular}
\caption{Summary of the main physical quantities characterizing the two-qubit--resonator system considered in this work. For each physical quantity in the left column, the right column provides the corresponding symbol and, where applicable, its defining equation.}
\label{table1}
\end{table}

\begin{figure*}[t]
\centering
\includegraphics[width=1\linewidth]{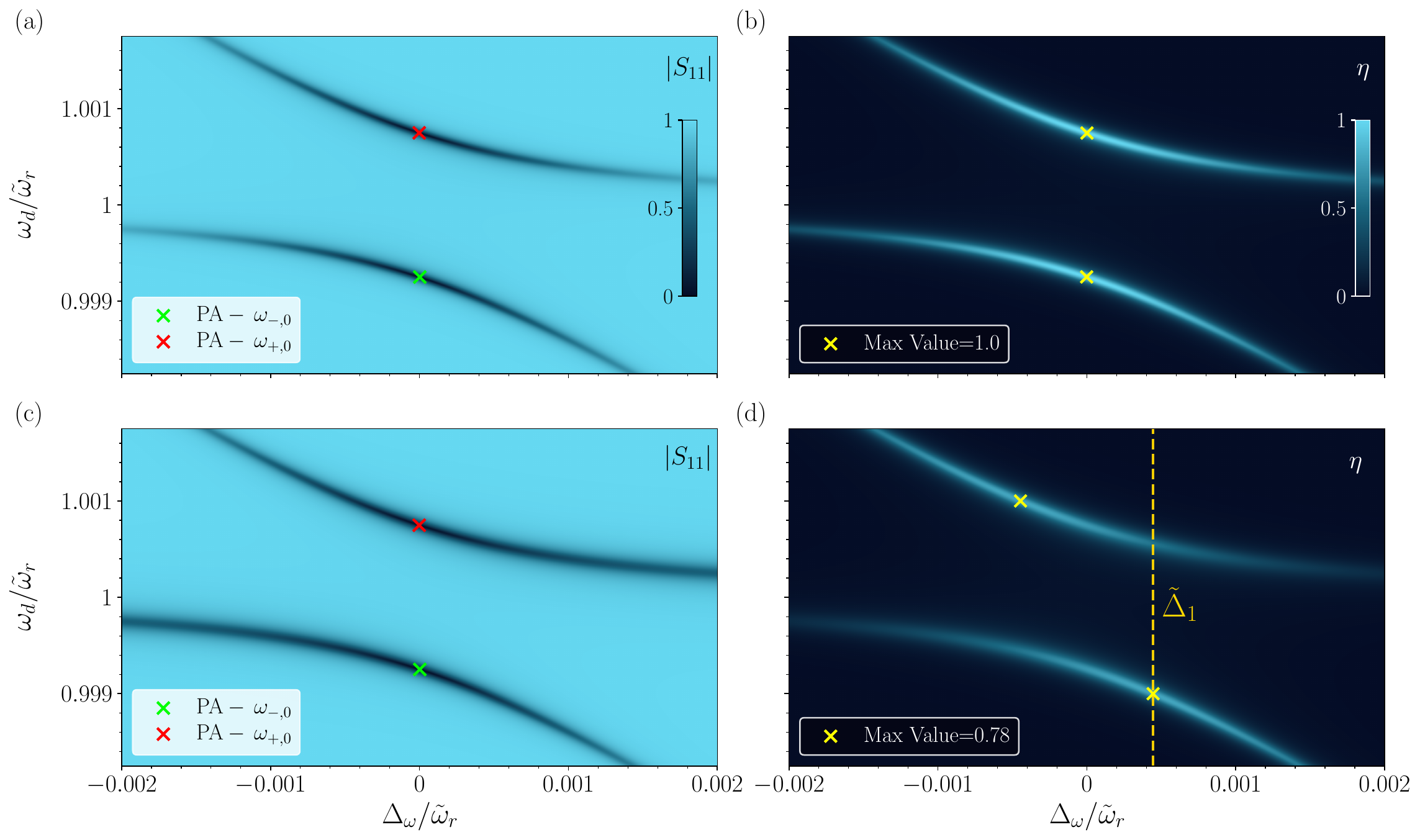}
\caption{Reflection and conversion efficiency with and without resonator intrinsic losses, for identical qubits and in the case of PT symmetry for $H_\mathrm{PA}$, i.e., under the loss-balancing condition. (a,\,c) Reflection $|S_{11}|$ and (b,\,d) conversion efficiency $\eta_j$ maps as a function of detuning $\Delta_\omega$ and driving frequency $\omega_d$, shown without (a,\,b) and with (c,\,d) non-radiative resonator losses. The crosses in the reflection maps mark the points of perfect absorption ($|{S_{11}}| = 0$), with red and green denoting the upper and lower hybrid modes, respectively. The yellow crosses in the efficiency maps denote the respective maxima. In (d), the vertical dashed line identifies the optimal detuning $\tilde{\Delta}_1$ maximizing $\eta_j$, and by symmetry, an analogous optimum exists at $-\tilde{\Delta}_1$ for the other hybrid mode. The parameters used are: $\gamma^{\rm qr}_1/\tilde{\omega}_{r} = \gamma^{\rm qr}_2/\tilde{\omega}_{r} = 6.25 \times 10^{-5}, \;g_{\rm eff}/\tilde{\omega}_r = 7.5 \times 10^{-4}, \; \gamma^{\rm qnr}_{1} = \gamma^{\rm qnr}_{2} = 0, \; \abs{A_d} = \gamma^\mathrm{r} / 100$; additionally, $\gamma^{\rm r} = \gamma^{\rm qr} = \gamma^{\rm qr}_1 + \gamma^{\rm qr}_2 ,\; \gamma^{\rm nr} = 0$ for (a,\,b), while $ \gamma^{\rm r}/\tilde{\omega}_r = 1.75 \times 10^{-4} ,\; \gamma^{\rm nr}/\tilde{\omega}_r = 0.5 \times 10^{-4}$ for (c,\,d).
}
\label{fig1}
\end{figure*}

In summary, the identification of PA conditions is straightforward within the associated non-Hermitian framework in the low-excitation regime. 
In the following subsections, we investigate the qubit emission properties by computing the emission spectra and cross-correlation functions via a Lindblad master-equation approach. Under the rotating-wave approximation, the coherent drive applied to the resonator is incorporated into the system Hamiltonian by adding the term  $H_d(t) = \sqrt{ \gamma^{\rm r}} (-iA_d \,a \, \exp(i\omega_dt)+\rm h.c.)$ to $H_{\rm eff}$ in \eqref{Heff}. 
Moving into the drive rotating frame, we obtain a time-independent Hamiltonian $H'_{\rm eff}$ by applying a suitable unitary transformation to $H_{\rm eff} + H_d(t)$ (see \appref{appB} for all details). A steady-state density matrix can then be numerically computed through \cite{breuer2002theory,gardiner2004quantum,walls2008quantum,lidar2020lecturenotestheoryopen}
\begin{equation}\label{mastereq}
\Dot{\rho} =- i \comm{H'_{\rm eff}}{\rho} + \mathcal{L}_s \rho\,,
\end{equation}
where $\mathcal{L}_s$ indicates the Liouvillian superoperator. All numerical simulations are performed using the \texttt{QuantumToolbox.jl} framework \cite{Mercurio2025quantumtoolboxjl}.

\subsection{PT symmetry for $H_{\mathrm{PA}}$}\label{resPT}

We first discuss the special case in which the effective Hamiltonian $H_\mathrm{PA}$ in \eqref{hrz_bare} displays PT symmetry at zero detuning $\left( \tilde{\omega}_r = 2\tilde{\omega}_q \right)$, i.e., when $ \gamma^{\rm r} =  \gamma^{\rm nr} +  \gamma^{\rm qr} + \gamma^{\rm qnr}$.

Figure~\ref{fig1}(a) shows the reflection map $\abs{S_{11}}$ as a function of the detuning $\Delta_\omega$ and drive frequency $\omega_d$, in absence of non-radiative losses ($ \gamma^{\rm nr} = \gamma^{\rm qnr}_{1} = \gamma^{\rm qnr}_{2} = 0$). 
Throughout this work, $\Delta_\omega$ is varied by tuning the qubit frequencies $\tilde \omega_{\rm q}$. This procedure can be readily implemented in circuit QED setups, for instance, by adjusting the flux offset $\epsilon$ of the flux qubits. For the sake of simplicity, in this effective model, we neglect any dependence of $g_{\rm eff}$ on $\Delta_\omega$, which however can be easily included if required.
By numerically computing the steady-state density matrix $\rho_{\rm ss}$ in the frame of the drive, we evaluate the reflection coefficient as $ \abs{S_{11}} = \abs{1- \sqrt{ \gamma^{\rm r}} \expec{a}_{\rm ss}/{{A_d}}}$, where $\expec{a}_{\rm ss} = \Tr{a \rho_{\rm ss}}$. 
In the weak-excitation regime $\bar n(\omega_d) \ll 1$, where  $\bar n(\omega_d)$ represents the mean number of resonator photons populated by the coherent drive for $g_{\rm eff} = 0$ (see \eqref{nph}), $S_{11}$ is independent of the excitation amplitude $A_d$.
The numerically calculated reflection coefficient $S_{11}(\omega_d)$ exhibits excellent agreement with the analytical expression in \eqref{s11}.

The avoided level crossing in \figref{fig1}(a) is a clear signature of the SC regime, arising from the coherent coupling between $\ket{g,g,1}$ and $\ket{e,e,0}$. The dips in the reflection spectra are in excellent agreement with the real parts of the eigenvalues of $H_{\rm PA}$, i.e., the complex eigenfrequencies $\tilde{\Omega}_\pm$ (see \appref{appA}). 
Since for $\Delta_\omega = 0$ the Hamiltonian $H_\mathrm{PA}$ is PT-symmetric, we simultaneously observe at resonance PA for the upper and lower hybrid modes (i.e., $\abs{S_{11}}=0$), as shown by the crosses in \figref{fig1}(a). 
As noted above, this feature directly follows from PT symmetry, as it implies that $\tilde{\gamma}_1 = \tilde{\gamma}_2 = 0$ simultaneously, ensuring a full real eigenspectrum. 

Exploiting PA enables perfect feeding in the quantum nonlinear optical process under study, which can be leveraged to maximize the emission signals extracted from the qubit output channels.
We compute the emission efficiency for the $j$-th qubit as
\begin{equation}\label{eff}
\eta_j (\omega_d, \Delta_\omega) = \frac{\Phi^{\rm q}_{j,{\rm out}}}{\Phi^{\rm r}_{{\rm in}}} = \frac{\gamma^{\rm qr}_j \expec{\sigma^+_j \sigma_j^-}_{\rm ss}}{\abs{A_d}^2}\,,
\end{equation}
where $\Phi^{\rm q}_{j,{\rm out}} = \expec{b_{j,{\rm out}}^{\dagger (\rm r)} b_{j,{\rm out}}^{(\rm r)}}_{\rm ss}$ and $\Phi^{\rm r}_{{\rm in}} = \expec{a_{\rm in}^{\dagger (\rm r)} a_{\rm in}^{(\rm r)}} = \abs{A_d}^2$ represent the $j$-th qubit output and the resonator input photon rates, respectively. Using the input-output relations in \eqref{inoutqs} and assuming zero-average qubit input signals $\expecsm{b_{j,{\rm in}}^{\dagger (\rm r)} b_{j,{\rm in}}^{(\rm r)}} = 0$, the output rate simplifies to $\Phi^{\rm q}_{j,{\rm out}} = \gamma^{\rm qr}_j \expec{\sigma^+_j \sigma_j^-}_{\rm ss}$. The efficiency $\eta_j$ thus quantifies the fraction of excitations transferred, on average, from the resonator input channel to the qubit output port.

In \figref{fig1}(b), we report $\eta_j$ for identical qubits (i.e., $\eta_1=\eta_2$) as a function of the drive frequency $\omega_d$ and detuning $\Delta_\omega$, using the same parameters adopted for the reflection spectra in \figref{fig1}(a). 
In the absence of non-radiative losses, the PA ensures $100 \%$ conversion efficiency at $\Delta_\omega = 0$ and at driving frequencies matching the PA condition, as marked by the yellow crosses in \figref{fig1}(b).
Furthermore, the coherent amplitude of each qubit output field vanishes, $\expecsm{b^{\rm (r)}_{j,{\rm out}}}_{\rm ss}=\sqrt{\gamma^{\rm qr}_j} \expecsm{\sigma^-_j}_{\rm ss}$, demonstrating that the emitted photons carry no first-order coherence, as in spontaneous down-conversion processes.

We now turn our attention to the impact of the non-radiative losses. Specifically, we introduce a finite resonator non-radiative loss $ \gamma^{\rm nr}$, while maintaining the loss-balance condition that ensures PT symmetry for $H_\mathrm{PA}$ at zero detuning, namely $\gamma^{\rm r} =  \gamma^{\rm nr} + \gamma^{\rm qr}$. Therefore, PA still occurs at zero detuning, as in \figref{fig1}(a), although the hybrid-mode linewidths slightly broaden [see \figref{fig1}(c)]. In this scenario, the efficiency $\eta_j$  in \figref{fig1}(d) decreases compared to \figref{fig1}(b) due to $ \gamma^{\rm nr} \neq 0$, and never reaches unity. Interestingly, the efficiency maxima no longer coincide with the PA points of $\abs{S_{11}}$, but instead appear at non-zero, symmetrically opposite detuning values $(\Delta_\omega/\tilde\omega_r = \pm \tilde{\Delta}_1)$, marked by yellow crosses in \figref{fig1}(d). At these detunings, the relevant hybrid mode (i.e., the one displaying the efficiency maximum) becomes more \textit{atom-like}, meaning that the corresponding eigenstate $\ket{\psi_{\pm}}$ in Eqs.~(\ref{eq4}) and (\ref{eq5}) exhibits a more pronounced contribution from the atomic state $\ket{e,e,0}$. This \textit{atom-like} character reduces of the effective resonator non-radiative loss seen by the $j$-th hybrid mode by a factor $|C_{j1}|^2 < 1/2$. Therefore, the optimal detunings, where the maxima of $\eta_j$ occur, result from the trade-off between PA and the reduced effective non-radiative loss at non-zero detunings.
In \appref{app_eff_PA}, we provide a rigorous analytical demonstration of how the non-radiative resonator losses $\gamma^\mathrm{nr}$ shift the efficiency maxima away from the PA condition, explaining the detuning discrepancy observed above.

The numerical results concerning the efficiency are confirmed analytically in \appref{appB}, where a steady-state solution of the truncated master equation yields the density matrix $\rho_{\rm ss}$ and the associated expectation values. For the lower hybrid mode, the analytical emission efficiency reads
\begin{equation}\label{emisslow}
\eta^{\rm th}_{j}(\omega_d) \approx \frac{2 \,\gamma^{\rm qr}_{j} \, \gamma^{\rm r}\,\sin^2{(\frac{\alpha}{2})}\cos^2{(\frac{\alpha}{2})}}{(\omega_{d}-\omega_{-,0})^2+(\frac{1}{2}\,\gamma_2)^2} \,,
\end{equation}
where $\gamma_{2}$ is the lower hybrid-mode loss rate, obtained by reversing the sign of $ \gamma^{\rm r}$ in the definition of $\tilde\gamma_j$ in \eqref{gammabar}. An analogous expression for the upper mode is reported in the same appendix. As illustrated by the comparative plots in \appref{appB}, namely \figref{fig10}, this analytical result is in good agreement with the exact numerical approach. At resonance (i.e., $\omega_d =\omega_{-,0}$), \eqref{emisslow} correctly peaks at the optimal positive detuning value indicated by the yellow cross in \figref{fig1}(d). Appendix \ref{app_eff_PA}, instead, focuses on the relation between the presence of non-radiative losses and the detuning corresponding to the maximum efficiency.
An analogous argument applies to the upper mode at the opposite detuning, which peaks at $\omega_d =\omega_{+,0}$. These results suggest that the efficiency could be further increased by achieving the PA condition at $\Delta_\omega \neq 0$. This possibility can be realized by exploiting Hermitian subspaces rather than PT symmetry, as investigated in \cref{Hrzsubsp}.

Further information on the emission processes is revealed by the steady-state spectrum of the $j$-th qubit, defined as
\begin{equation}\label{emiss.spec}
\tilde{S}_{{\rm q},j}(\omega)= 2\Re{\int^\infty_0 d\tau\, e^{-i\omega \tau} \expec{{\sigma}_j ^{+}(\tau) \sigma^{-}_j(0)}_{\rm ss}}\,,
\end{equation}
where $\omega$ denotes the emission frequency, implicitly including a $\omega_d/2$ shift to account for the transformation to the rotating frame of the drive.
As expected, it is found that the emission spectrum shows more pronounced peaks for parameters corresponding to the maximum emission efficiency (yellow crosses in \figref{fig1}). 

\begin{figure}[t]
    \centering
    \includegraphics[width=1\linewidth]{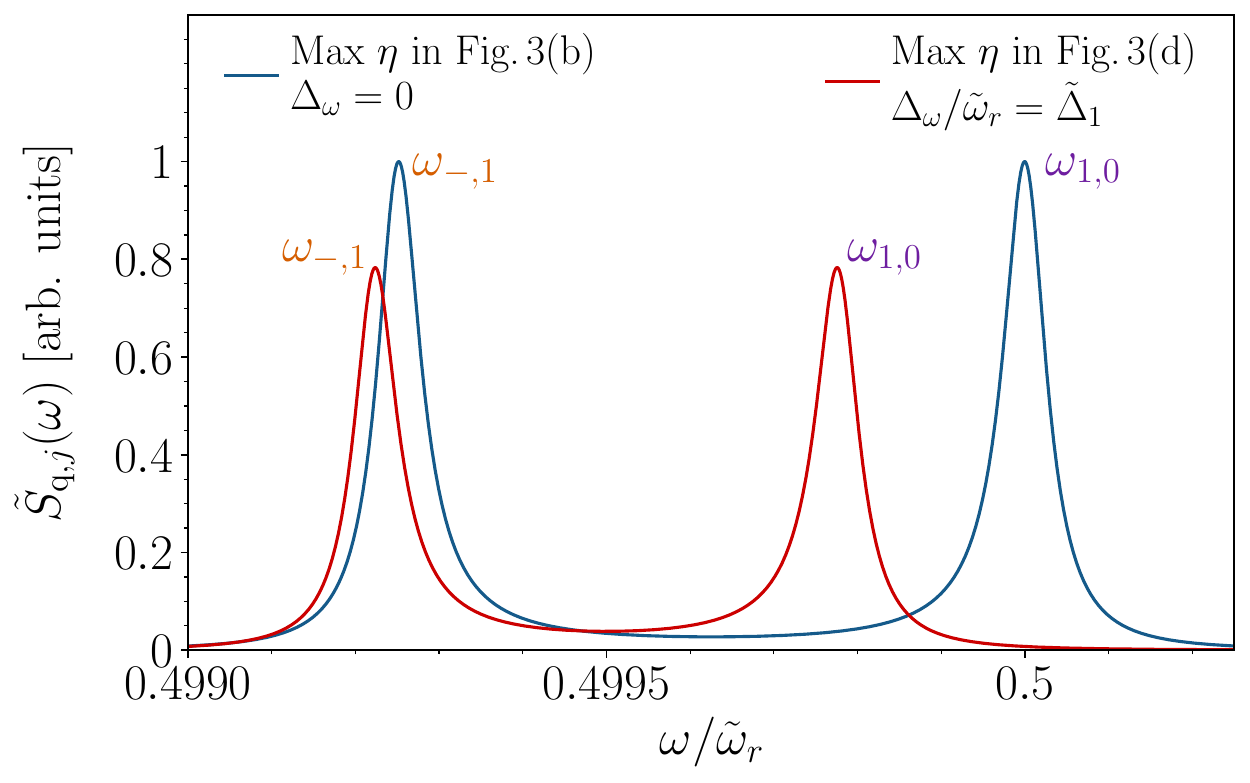}
    \caption{
    Emission spectra $\tilde{S}_{\mathrm{q},j}$ for the qubit output channels, 
for two identical qubits (i.e., $\tilde{S}_{\mathrm{q},1}=\tilde{S}_{\mathrm{q},2}$), 
normalized to their global maximum. The spectral peaks reflect the cascaded 
spontaneous down-conversion process and correspond to the energy-level 
transitions sketched in \figref{fig_avoided_crossing}(c). Each curve is 
evaluated at the drive frequency $\omega_d$ and detuning $\Delta_\omega$ 
that maximize the conversion efficiency $\eta_j$ for the lower hybrid 
mode in \figref{fig1}: the blue curve corresponds to the case without non-radiative losses (parameters of \figref{fig1}(b), $\Delta_\omega = 0$), while the red curve includes non-radiative resonator loss (parameters of \figref{fig1}(d), 
$\Delta_\omega/\tilde{\omega}_r = \tilde{\Delta}_1$).
    }
    \label{fig3_spettri}
\end{figure}

In \figref{fig3_spettri}, we plot the emission spectrum for identical qubits at the points of maximum efficiency for the upper hybrid mode, in the presence (red) and absence (blue) of non-radiative losses. 
Both spectra clearly exhibit the cascaded transitions associated with the level structure in \figref{fig_avoided_crossing}(c), with the two peaks in each plot matching the spontaneous-emission processes indicated by the orange $(\omega_{-,1})$ and purple $(\omega_{1,0})$ arrows.
In the presence of non-radiative losses, the peaks of the red curve are shifted relative to the blue one (since the efficiency maximum occurs at $\Delta_\omega = \tilde{\Delta}_1 \neq 0$) and display reduced intensity. Analogous results hold when the drive is tuned to match the efficiency maximum of the lower hybrid mode. This correlated pair emission, extracted directly from the qubit output channels, represents a spontaneous down-conversion process with applications ranging from quantum metrology and communication to fundamental tests of quantum mechanics \cite{CouteauContmpPhys2018, CouteauNatureRev2023}.

\begin{figure}[b]
    \centering
    \includegraphics[width=1\linewidth]{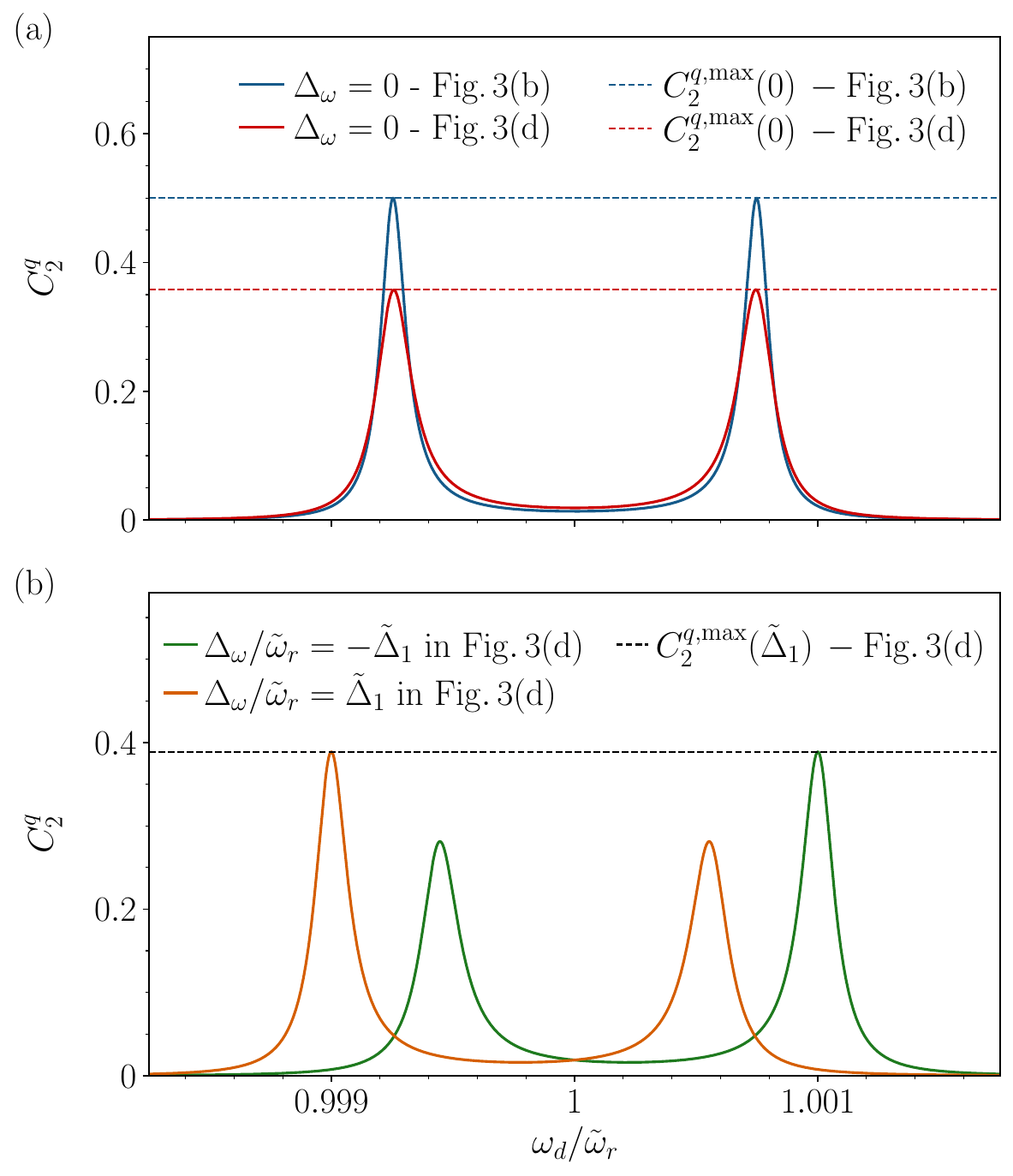}
    \caption{
    Two-qubit output correlation function $C_2^q$ as a function of the drive frequency $\omega_d$ for fixed detuning values $\Delta_\omega$ (see legends). The blue curve in panel (a) is computed adopting the parameters from \figref{fig1}(b), while the red curve in (a) and the curves in panel (b) are obtained using those from \figref{fig1}(d). Horizontal lines indicate the maximum achievable correlation $C_2^{q,\max}(\Delta_\omega)$ at the hybrid-mode resonances.
    }
    \label{fig3}
\end{figure}

Another important property to analyze is the correlations between the two qubit output signals. To this end, we compute the equal-time steady-state normalized two-qubit output correlation function
\begin{equation}\label{corr}
C^q_2 = \sqrt{\gamma^{\rm qr}_1 \gamma^{\rm qr}_2} \frac{\expec{\sigma^+_1 \sigma_1^- \sigma^+_2 \sigma_2^-}_{\rm ss}}{|A_d|^2}\,.
\end{equation}
In a continuous-wave driven-dissipative configuration, this quantity satisfies the inequality $C^q_2 \le C^{q,\mathrm{max}}_2$, with the upper bound defined by 
\begin{equation}\label{c2max}
    C^{q,\mathrm{max}}_2 (\Delta_\omega) = \frac{\sqrt{\eta^{\rm res}_1 (\Delta_\omega) \, \eta^{\rm res}_2 (\Delta_\omega)}}{2} \, ,
\end{equation}
where $\eta^{\rm res}_j (\Delta_\omega)$ denotes the maximum value of the conversion efficiency $\eta_j$ at the specified detuning, obtained when $\omega_d$ is tuned to resonance with the relevant hybrid mode.
The $1/2$ factor reflects the competition between coherent pair generation and independent spontaneous emission of the two qubits. Since the individual decay events remain temporally uncorrelated, the equal-time coincidence rate is suppressed relative to the product of the individual emission rates \cite{walls2008quantum}.
For identical qubits ($\eta_1 = \eta_2 = \eta$), the maximum correlation reduces to $C^{q,\mathrm{max}}_2 = \eta^\mathrm{res} / 2$.

Figure \ref{fig3} shows $C^q_2$ as a function of $\omega_d$, both in the presence and absence of resonator non-radiative losses. In particular, \figref{fig3}(a) refers to the case of zero detuning $\Delta_\omega = 0$, i.e., in correspondence of PA, for $\gamma^\mathrm{nr} = 0$ (blue) and $\gamma^\mathrm{nr} \neq 0$ (red), with the loss-balance condition $\gamma^{\rm r} =  \gamma^{\rm nr}+ \gamma^{\rm qr}$ preserved in both cases.
At this detuning, both curves reach the maximum achievable correlation $C^{q,\mathrm{max}}_2 (0)$ at the hybrid mode frequencies, with equal peak heights, while the presence of non-radiative losses only reduces the overall magnitude of the peaks. 

Figure \ref{fig3}(b) shows that, as expected, non-radiative losses shift the optimal working point away from PA: the maximum correlation $C^{q,\mathrm{max}}_2$ is achieved at the detuning corresponding to peak efficiency $\tilde{\Delta}_1$, only for the peak corresponding to the relative hybrid mode (yellow crosses in \figref{fig1}(d)), with the complementary behavior appearing at the opposite detuning for the other hybrid mode. We also find (not shown) that the bound $C^{q,\mathrm{max}}_2 = \sqrt{\eta^{\rm res}_1 \, \eta^{\rm res}_2}/2$ is saturated only when $\gamma^{\rm qr}_1 = \gamma^{\rm qr}_2$: an imbalance in the radiative decay rates produces a mismatch in the qubit emission profiles that prevents the system from reaching maximal correlation. The curves in \figref{fig3} are in good agreement with the analytical model derived in \appref{appB} (see \figref{fig10}(b)).

It is worth noting that the population correlation $\expec{\sigma^+_1 \sigma_1^- \sigma^+_2 \sigma_2^-}_{\rm ss}$ in the numerator of $C^{q}_2$ does not oscillate at $\omega_d$ and is therefore susceptible to thermal noise. From an experimental perspective, this poses a challenge, and it would be more advantageous to access $\abs{\langle \sigma_1^- \sigma_2^- \rangle_{\rm ss}}^2$, which is closely linked to its coherent part. Indeed, since the correlation $\langle \sigma_1^- \sigma_2^- \rangle_{\rm ss}$ oscillates at the driving frequency $\omega_d$ with a phase locked to that of the coherent input tone, it is intrinsically robust against the thermal background. For example, this two-qubit correlation could be measured by taking the Fourier transform at frequency $\omega_d$ of the cross-correlation of the output voltage signals, $\mathcal{F}[\langle V^X_{1,{\rm out}}(t)V^X_{2,{\rm out}}(t) \rangle]$, where the output voltage $V^X_{j,{\rm out}}(t)$ is proportional to the field quadrature $X_j \propto \left(\sigma_j^- e^{i\omega_d t/2} + \text{H.c.}\right)$, with $\langle V^X_{j, {\rm out}} \rangle = 0$. This procedure yields a signal directly proportional to the real part of $\langle \sigma_1^- \sigma_2^- \rangle_{\rm ss}$, as shown theoretically in Ref.~\cite{orlando2026quantumvacuumradiation}, albeit for a different system. Furthermore, the imaginary part of $\langle \sigma_1^- \sigma_2^- \rangle_{\rm ss}$ can be extracted by measuring the cross-correlation between two different quadratures, for instance, $V^X_{1,{\rm out}}$ and $V^Y_{2,{\rm out}} \propto Y_2$, obtained by phase-shifting the local oscillator of the second channel by $\pi/2$. Together, these two measurements enable the full reconstruction of $\abs{\langle \sigma_1^- \sigma_2^- \rangle_{\rm ss}}^2$.

\begin{figure*}[htb]
    \centering
    \includegraphics[width=\linewidth]{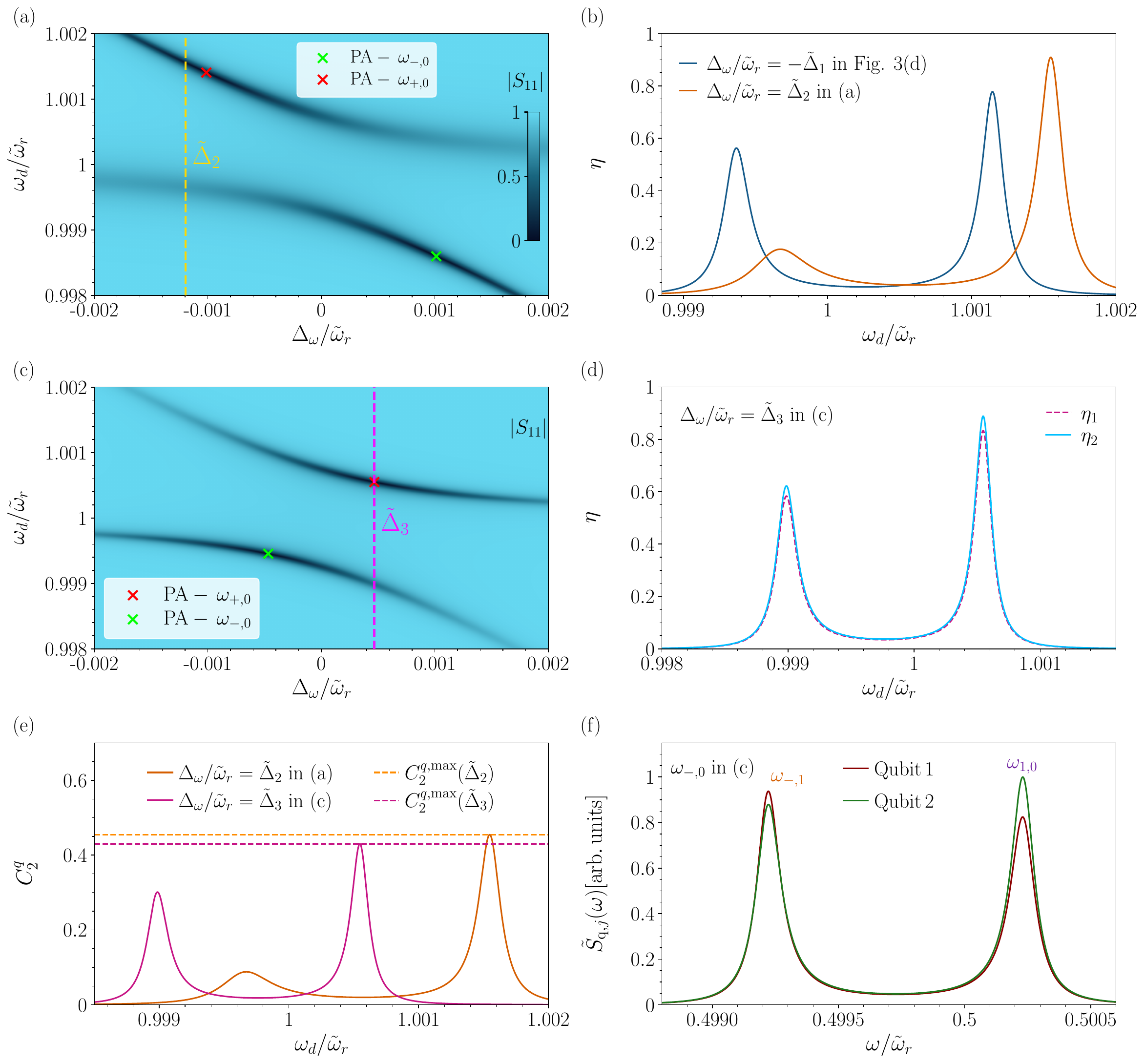}
    \caption{
    (a,\,c) Reflection coefficient $\abs{S_{11}}$ as a function of detuning $\Delta_\omega$ and drive frequency $\omega_d$, under broken loss balance ($\gamma^{\rm r}\neq\gamma^{\rm nr}+\gamma^{\rm qr}+\gamma^{\rm qnr}$), with non-radiative losses in the resonator (a) or qubits (c). Green and red crosses mark the PA points ($\abs{S_{11}}=0$) for the lower and upper hybrid modes, respectively, where a Hermitian subspace of $H_{\rm PA}$ emerges. Vertical dashed lines identify the optimal detuning $\tilde{\Delta}_j$ maximizing the conversion efficiency of the respective hybrid mode. 
    (b) Conversion efficiency $\eta_j$ as a function of $\omega_d$ for identical qubits ($\eta_1=\eta_2$) at fixed detunings (see legend). The broken-loss-balance curve (orange) is compared with the PT-symmetric case of \figref{fig1}(d) (blue). 
    (d) Individual efficiencies $\eta_1$ and $\eta_2$ as functions of $\omega_d$ for qubits with different non-radiative losses. 
    (e) Two-qubit output correlation $C^q_2$ as a function of $\omega_d$ for the two configurations in (a) and (c), at the respective optimal detunings (see legends). Horizontal lines indicate the maximum achievable correlation $C_2^{q,\max}(\tilde{\Delta}_j)$, with the peaks located at the corresponding hybrid-mode resonance.
    (f) Qubit emission spectra $\tilde{S}_{{\rm q},j}(\omega)$ at the PA point marked by the green cross in (c), normalized to the global maximum. The two peaks correspond to the labeled cascade transitions of \figref{fig_avoided_crossing}(c) (orange and purple arrows). 
    Plots in (a) and the orange curves in (b,\,e) use the parameters of \figref{fig1}(d), except $\gamma^{\rm r}/\tilde{\omega}_r = 5\times10^{-4}$. Plots in (c,\,d), the purple curve in (e), and the spectra in (f) are calculated with $\gamma^{\rm r}/\tilde{\omega}_r = 1.25\times10^{-4}$, $\gamma^{\rm qr}_{1,2}/\tilde{\omega}_r = 1\times10^{-4}$, $g_{\rm eff}/\tilde{\omega}_r = 7.5\times10^{-4}$, $\gamma^{\rm nr}=0$, $\gamma^{\rm qnr}_1/\tilde{\omega}_r = 2\times10^{-5}$, $\gamma^{\rm qnr}_2/\tilde{\omega}_r = 1.25\times10^{-5}$, and $|A_d|=\gamma^{\rm r}/50$.
    }
    \label{fig4}
\end{figure*}

To gain further insight into the spectral properties of the emitted correlations, complementing the equal-time analysis of $C^q_2$ carried out above, we define the cross-correlation spectrum as \cite{Gasparinetti2017}
\begin{equation}\label{crossex}
    {\cal S}_{j,k}(\omega) =  2\,\gamma^{\rm qr}_{j}\Re{\int^\infty_0 d\tau\, e^{-i\omega \tau} \expec{{P}^e_j (\tau) P^e_k(0)}_{\rm ss}}\,,
\end{equation}
where $j \neq k$ and $P^e_j = \sigma^{+}_{j}\sigma^{-}_{j} = (|e\rangle\langle e|)_j$ denotes the projector onto the excited state of the $j$-th qubit. For identical qubits ($\gamma^{\rm qr}_1 = \gamma^{\rm qr}_2$), the spectrum is symmetric, ${\mathcal S}_{1,2}(0) = {\mathcal S}_{2,1}(0)$, as a direct consequence of the symmetry of the dissipative channels within the Liouvillian framework.
We find that ${\cal S}_{1,2}(0)$ is maximized at the same parameter values $(\Delta_\omega, \omega_d)$ that maximize the emission efficiency $\eta_j$ (yellow crosses in \figref{fig1}). At these points, in the absence of non-radiative losses, the zero-frequency limit reaches the maximum quantum limit for pair correlations \cite{walls2008quantum}:
\begin{equation}\label{P1n}
    P_{12}= \sqrt{\langle P^e_1 \rangle_{\rm ss} \langle P^e_2 \rangle_{\rm ss}}\,,
\end{equation}
where the steady-state populations $\langle P^e_j \rangle_{\rm ss}$ are evaluated at the same $\Delta_\omega$ and $\omega_d$. This condition marks a regime of maximal spectral cross-correlation and carries a direct physical interpretation: the photons emitted by the two qubits can be detected as simultaneous pairs within their collective emission window, representing a signature of a highly non-classical process.
The introduction of non-radiative losses ($\gamma^{\rm nr} \neq 0$) breaks the equality ${\mathcal S}_{1,2}(0) = P_{12}$, leading to a reduction of spectral cross-correlations. 
Furthermore, the condition for maximal spectral cross-correlation can be met only for symmetric decay rates, $\gamma^{\rm qr}_1 = \gamma^{\rm qr}_2$.
This is consistent with the well-known degradation of intensity correlations in twin-beam sources caused by losses beyond the radiative decay channel \cite{BondaniPRA2007, walls2008quantum, LoseroSciRep2018, KopylovPRR2025}, in analogy with the theoretical framework and experimental observations of twin-beam generation \cite{BoyerSCI2008, BridaNatPhot2010, PontulaPRXQM2024}. This behavior will be further explored across the parameter space in the following subsection.

\subsection{Hermitian Subspaces for $H_{\mathrm{PA}}$}\label{Hrzsubsp}

The results of the previous subsection expose an intrinsic limitation of the PT-symmetric configuration: once non-radiative losses are present, the loss-balance condition required for PT symmetry pins PA to zero detuning, while the actual efficiency maximum is pulled away from it, namely toward the detuning at which the relevant hybrid mode becomes more atom-like. Since non-radiative losses are generally unavoidable in realistic implementations, and loss balance is itself an additional, non-trivial requirement on the system parameters, this raises a natural question: can PA be relocated closer to the efficiency maximum without relying on any loss-balance condition at all? 
We show that this is indeed possible by relaxing the PT-symmetry requirement and instead exploiting the emergence of Hermitian subspaces within $H_{\mathrm{PA}}$. As we discuss below, this strategy provides a systematic route to enhance the conversion efficiency, even in the presence of significant, unbalanced non-radiative losses.

To this end, we consider a configuration in which the loss-balance condition is deliberately broken. Figure~\ref{fig4}(a) shows the reflection spectra as a function of $\Delta_\omega$, adopting (for a better comparison) the same $g_{\rm eff}$, $\gamma^{\rm qr}_1 = \gamma^{\rm qr}_2$, and $\gamma^{\rm nr}$ as in \figref{fig1}(c), but now with a non-radiative resonator loss rate $\gamma^{\rm nr}$ so that $\gamma^{\rm r} \neq \gamma^{\rm nr} + \gamma^{\rm qr}$, i.e., the PT-symmetry condition for $H_{\mathrm{PA}}$ no longer holds. The avoided crossing between the upper and lower hybrid modes, typical of the SC regime, is still clearly visible upon varying $\Delta_\omega$. Despite the absence of PT symmetry, however, PA is not lost, as it reappears on the upper (red cross) or lower (green cross) hybrid-mode branch at opposite, finite values of $\Delta_\omega$. This is precisely the signature of a Hermitian subspace of $H_{\mathrm{PA}}$: with the loss rates fixed, the detuning acts as an external knob that, through the Hopfield coefficients $|C_{jk}|^2$ entering via the mixing angle $\alpha$, reshapes the light-matter composition of the hybrid modes until the effective loss $\tilde{\gamma}_j$ of one of them vanishes, as previously discussed (see \eqref{Hrz} and \eqref{gammabar}). When this happens, the corresponding eigenvalue of $H_{\mathrm{PA}}$ becomes purely real, and $\abs{S_{11}}$ vanishes at $\omega_d = \tilde{\Omega}_j$, reproducing, without any loss-matching requirement, the same reflection zeros observed in \secref{resPT} under PT symmetry. These results are consistent with the experimental findings of Ref.~\cite{BonizzoniNatureComm2025}.

The natural question, then, is whether this relocation of PA translates into an efficiency gain. Figure~\ref{fig4}(b) answers this by comparing the qubit emission efficiency $\eta_j$ ($\eta_1=\eta_2$) for this configuration (orange curve) with the PT-symmetric case of \figref{fig1}(d) (blue curve), in both cases evaluated at the $\Delta_\omega$ that maximizes $\eta_j$ (namely $-\tilde{\Delta}_1$ and $\tilde{\Delta}_2$, respectively). A clear enhancement of conversion efficiency is observed: at the hybrid-mode resonance, the maximum $\eta_j$ increases from approximately $0.77$ to about $0.90$. Crucially, this strategy circumvents the strict loss-matching requirement of PT symmetry, allowing the system to operate even in regimes where $ \gamma^{\rm r} \gg  \gamma^{\rm nr}$, while the introduced detuning still recovers near-perfect feeding. As in \secref{resPT}, the optimal detuning $\tilde{\Delta}_2$ that maximizes $\eta_j$ (vertical yellow dashed line in \figref{fig4}(a)) lies slightly off the exact PA point (red cross), although here this discrepancy is considerably smaller.  
Physically, as already discussed in the PT-symmetric case, this improvement can be traced to the increasingly \textit{atom-like} character of the hybrid mode, which suppresses the effective non-radiative resonator decay rate ($\gamma^{\rm nr}|C_{j1}|^2$) while still allowing efficient power transfer from the input resonator to the two-qubit excitations.
The same picture holds for the opposite detuning, $-\tilde{\Delta}_2$, where efficiency enhancement occurs at the resonance of the other hybrid mode.

So far, non-radiative losses have been confined to the resonator. We now consider the complementary scenario, in which non-radiative losses instead act (asymmetrically) only on the qubits ($\gamma^{\rm qnr}_1 \neq \gamma^{\rm qnr}_2 \neq 0$), setting $\gamma^{\rm nr}=0$, still under broken loss balance ($\gamma^{\rm r} \neq \gamma^{\rm nr} + \gamma^{\rm qr} + \gamma^{\rm qnr}$). 
Figure~\ref{fig4}(c) shows the corresponding reflection map. Fixing $\Delta_\omega$ at the PA value marked by the red cross, we plot the individual efficiencies $\eta_1$ and $\eta_2$ in \figref{fig4}(d). Here, unlike the previous case, the optimal detuning $\tilde{\Delta}_3$ (vertical purple dashed line) coincides exactly with the corresponding PA point, so that the efficiency maximum and the reflection zero overlap with no discrepancy in the $(\omega_d, \Delta_\omega)$ map. As we show formally in \appref{app_eff_PA}, this exact correspondence follows directly from $\gamma^{\rm nr}=0$. Specifically, the detuning maximizing $\eta_j$ and the detuning at which $\tilde\gamma_j$ vanishes are generally distinct whenever the resonator carries a non-radiative loss, and coincide only in its absence, regardless of how large or asymmetric the qubit non-radiative losses are. We note that $\eta_1 < \eta_2$ when $\omega_d$ is resonant with the relevant (upper) hybrid-mode frequency, an asymmetry directly traceable to the choice $\gamma^{\rm qnr}_1 > \gamma^{\rm qnr}_2$. Remarkably, despite this non-radiative matter dissipation, the system still reaches peak efficiencies of approximately $0.90$ for $\eta_2$ and $0.80$ for $\eta_1$. In contrast to the case discussed above, the improvement here stems from the increasingly \textit{photon-like} character of the hybrid modes, which suppresses the effective non-radiative matter losses ($\gamma^{\rm qnr}\abs{C_{j2}}^2$).

Having established the efficiency gains associated with the Hermitian subspaces of $H_{\rm PA}$, a natural question is whether this improvement also propagates to the two-qubit output correlations discussed in \secref{resPT}. Figure~\ref{fig4}(e) shows $C^{q}_2$ as a function of $\omega_d$ for the two configurations of panels (a) and (c), evaluated at their respective optimal detunings $\tilde{\Delta}_2$ and $\tilde{\Delta}_3$ (orange and purple curves, respectively). As in \figref{fig3}(b), both curves saturate the bound $C_2^{q,\rm max}(\tilde\Delta_j)$ (see \cref{c2max}) at the corresponding hybrid-mode resonance, confirming that maximal instantaneous correlation of the qubit output channels persists away from strict PT symmetry, with both symmetric and asymmetric qubit losses. 
Moreover, since $\eta_1^{\rm res}$ and $\eta_2^{\rm res}$ are themselves enhanced by the Hermitian-subspace mechanism, the peak values of $C^q_2$ in \figref{fig4}(e) exceed those obtained in \figref{fig3}(a,b), hence the efficiency gain discussed above translates directly into a stronger two-qubit correlation. This equal-time enhancement, however, does not extend to the cross-correlation spectrum. For the configuration of \figref{fig4}(a), the cross-correlation ${\mathcal S}_{1,2}(0)$, or equivalently ${\mathcal S}_{2,1}(0)$, does not reach the ideal quantum limit ${\mathcal S}_{1,2}(0) = P_{12}$, as a consequence of the combined effect of $\gamma^{\rm nr}\neq0$ and the broken loss-balance condition ($\gamma^{\rm r}\neq\gamma^{\rm nr}+\gamma^{\rm qr}$). 
Moreover, for the asymmetric-loss configuration of \figref{fig4}(c,d), ${\mathcal S}_{1,2}(\omega)$ and ${\mathcal S}_{2,1}(\omega)$ are no longer equal to each other (not shown here), due to the unequal non-radiative qubits' decay rates, which further prevents the system from reaching ${\mathcal S}_{j,k}(0)=P_{12}$, with larger deviations for increasing non-radiative decay rate.

To complete the picture, \figref{fig4}(f) shows the qubits' emission spectra defined in \eqref{emiss.spec}, evaluated at the drive frequency and detuning marked by the green cross in \figref{fig4}(c). As in \secref{resPT}, a photon-cascade process is again observed: the two peaks reproduce the spontaneous down-conversion transitions sketched in \figref{fig_avoided_crossing}(c), namely those marked by the orange ($\omega_{-,1}$) and purple ($\omega_{1,0}$) arrows, with spectral amplitudes that depend on the chosen qubit output channel.

\begin{figure*}[t]
    \centering
    \includegraphics[width=\linewidth]{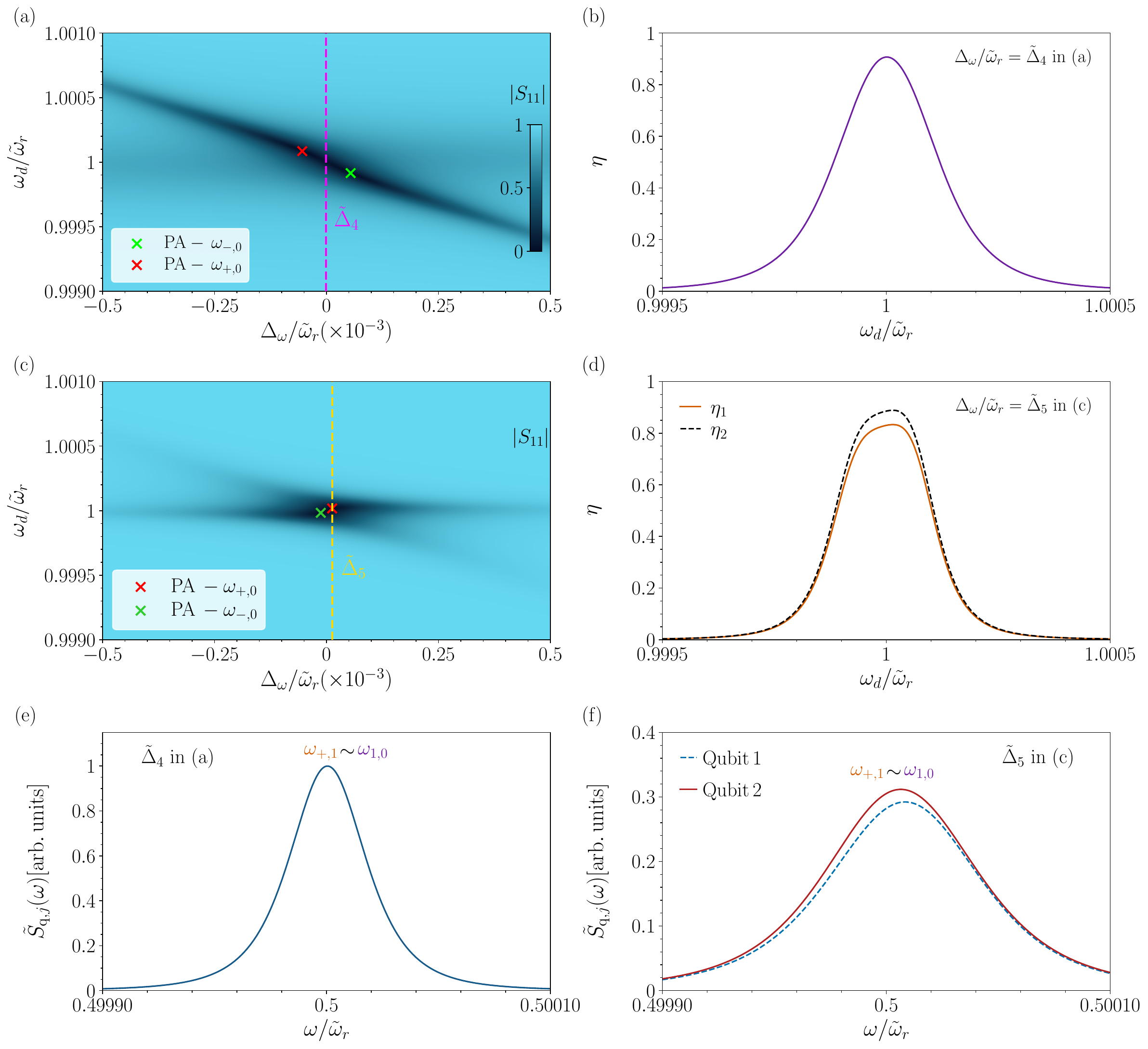}
     \caption{(a,\,c) Reflection coefficient $\abs{S_{11}}$ as a function of detuning $\Delta_\omega$ and drive frequency $\omega_d$, under broken loss balance ($\gamma^{\rm r}\neq\gamma^{\rm nr}+\gamma^{\rm qr}+\gamma^{\rm qnr}$) and in the weak nonlinear coupling regime ($g_{\rm min} < g_{\rm eff} < g_{\rm th}$), with non-radiative losses in the resonator (a) or qubits (c). The crosses mark the PA points ($\abs{S_{11}}=0$), while the vertical dashed lines identify the optimal detuning $\tilde{\Delta}_j$ maximizing the conversion efficiency of the respective hybrid mode. 
    (b) Conversion efficiency $\eta_j$ ($\eta_1=\eta_2$) as a function of $\omega_d$ at fixed $\Delta_\omega$ (see legend). 
    (d) Individual efficiencies $\eta_1$ and $\eta_2$ as functions of $\omega_d$ for qubits with different non-radiative losses. 
    (e,\,f) Qubit emission spectra $\tilde{S}_{{\rm q},j}(\omega)$, normalized to their global maximum, evaluated at the maximum efficiency points in (a,\,c), respectively. The labeled cascade transitions of \figref{fig_avoided_crossing}(c) are no longer sufficiently resolved, merging into a single emission peak. 
    Parameters in~(a,\,b,\,e) are the same as in \figref{fig4}(a\,,b), except $g_{\rm eff}/\tilde{\omega}_r = 1.25\times10^{-4}$. Analogously, parameters in~(c,\,d,\,f) are the same as in \figref{fig4}(c\,,d), except $g_{\rm eff}/\tilde{\omega}_r = 8.75\times10^{-5}$.}
    \label{fig5}
\end{figure*}

We remark that all findings in this subsection have been obtained in the weak-excitation limit, $\bar{n}(\omega_d) \ll 1$, and under the assumption of identical dressed qubit frequencies ($\tilde{\omega}_{{\rm q},1} = \tilde{\omega}_{{\rm q},2}$). As detailed in \appref{appD}, however, this second assumption is not essential: all results extend straightforwardly to the non-degenerate case ($\tilde{\omega}_{{\rm q},1} \neq \tilde{\omega}_{{\rm q},2}$), leaving both the PA condition and the conversion efficiency unaffected. The only difference resides in the transition frequencies appearing in the emission spectra $\tilde{S}_{q,j}$, which reflect the richer level structure of \figref{fig_avoided_crossing}(b).

The analysis carried out so far has focused on the SC regime ($g_{\rm eff} \gg \gamma^{\rm r},\, \gamma^{\rm nr},\, \gamma^{\rm qr}\,,\,\gamma^{\rm qnr}$). As discussed in \secref{onephtwatoms}, this condition is necessary to guarantee an efficient and reversible quantum state transfer in the time domain between $|g,g,1\rangle$ and $|e,e,0\rangle$. From a spectral standpoint, this manifests as a well-resolved avoided crossing in $\abs{S_{11}}$, with the two hybrid-mode branches clearly separated. 
However, reaching such a strong effective coupling remains experimentally challenging, as it has so far been demonstrated only in a few circuit-QED implementations \cite{WangNatureComm2025, TomonagaNatureComm2025}, in the absence of strong driving fields. This naturally motivates us to assess whether our PA-based optimization strategy retains its effectiveness in the weak nonlinear regime.


\subsection{Weak effective coupling}\label{weakeffcoupl}

We now extend the PA-based optimization strategy of \secref{Hrzsubsp} to the weak nonlinear coupling regime.
The strong-to-weak transition threshold is traditionally defined as
\begin{equation}\label{gth}
    g_{\rm th}= \frac{ \gamma^{\rm r}+  \gamma^{\rm nr}+   \gamma^{\rm qr} + \gamma^{\rm qnr}}{4}\,.
\end{equation}
As discussed in Refs.~\cite{zanottoSCIREP2016,BonizzoniNatureComm2025}, PA does not disappear immediately below $g_{\rm th}$. On the contrary, it still persists as long as $g_{\rm eff}$ remains above the minimum value
\begin{equation}\label{gmin}
    g_{\rm min} = \frac{1}{2}\sqrt{(\gamma^{\rm qr} + \gamma^{\rm qnr})\,( \gamma^{\rm r}- \gamma^{\rm nr})}\,.
\end{equation}
Throughout this section, we explore the weak nonlinear coupling regime with the presence of PA, i.e., \\$g_{\rm min}<g_{\rm eff}<g_{\rm th}$.

Figure~\ref{fig5}(a) shows the reflection spectra $\abs{S_{11}(\omega_d)}$ for a scenario analogous to \figref{fig4}(a) with the only difference of a smaller $g_{\rm eff}$, where the PA points are marked by the red and green crosses.
Despite the weak coupling and the presence of resonator non-radiative loss ($\gamma^{\rm nr}\neq0$), the efficiency $\eta_j$ ($\eta_1=\eta_2$) still reaches values of approximately $0.90$, comparable to its SC counterpart in \figref{fig4}(b) (orange curve), as shown in \figref{fig5}(b). Even in this case, due to the presence of non-radiative resonator losses (see \appref{app_eff_PA}), the optimal detuning $\tilde{\Delta}_4$ maximizing $\eta_j$ for the upper hybrid mode (vertical purple dashed line) lies slightly off the exact PA point.

The same robustness holds when, instead, asymmetric non-radiative qubit losses ($\gamma^{\rm qnr}_1\neq\gamma^{\rm qnr}_2$) are introduced with $\gamma^{\rm nr}=0$, mirroring the configuration and parameters of \figref{fig4}(c,\,d), albeit with a smaller coupling. Figure~\ref{fig5}(c) shows the resulting reflection map, while \figref{fig5}(d) confirms that the efficiency maxima remain nearly as high as in the SC regime ($\eta_j\approx0.90$, to be compared to \figref{fig4}(d)), with the higher value naturally associated with the qubit experiencing the lower non-radiative dissipation $\gamma^{\rm qnr}_j$. Consistently with the discussion of \appref{app_eff_PA}, $\gamma^{\rm nr}=0$ again guarantees an exact coincidence between the efficiency maximum and the PA point, marked respectively by the vertical yellow dashed line ($\tilde\Delta_5$) and the red cross in \figref{fig5}(c).

To complete the picture, \figref{fig5}(e,\,f) show the qubits' emission spectra $\tilde{S}_{{\rm q},j}$ corresponding to panels (a) and (c). Unlike the SC case, the previously distinct cascade frequencies are now indistinguishable, merging into a single visible peak (associated with the two transitions identified by the orange, $\omega_{+,1}$, and purple, $\omega_{1,0}$, arrows in \figref{fig_avoided_crossing}(c)), whose amplitude depends on the chosen qubit output channel.

Having established that near-unity conversion efficiencies persist even in the weak-coupling limit, we now broaden the scope of this analysis. Specifically, in the following section, we map the dependence of the efficiency on the full set of circuit parameters using phase diagrams.

\section{Phase diagrams for the one-photon two-qubit process}\label{phasediagrams}

\begin{figure}[!h]
    \centering
    \includegraphics[width=\linewidth]{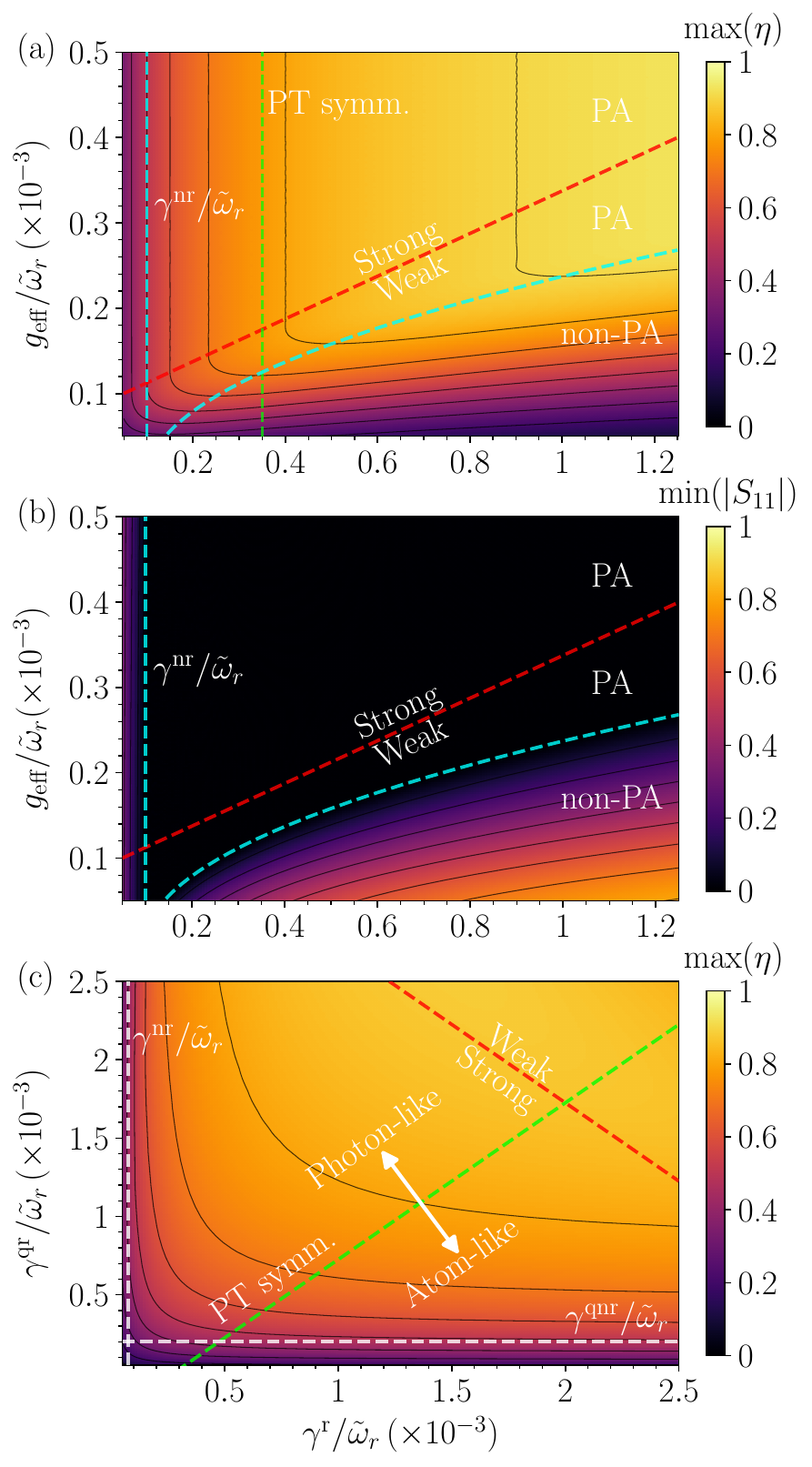}
    \caption{(a) Maximum efficiency $\max(\eta_j)$ and (b) reflection coefficient $|S_{11}|$ as functions of $\gamma^{\rm r}$ and $g_{\rm eff}$, for identical qubits. Each point of these diagrams is obtained by jointly optimizing over the drive frequency $\omega_d$ and the detuning $\Delta_\omega$. The vertical cyan dashed line marks the resonator non-radiative loss and, together with the curved cyan line, delimits the PA region ($g_{\rm eff}=g_{\rm min}$). The green line identifies the PT-symmetry condition for $H_{\rm PA}$ ($\gamma^{\rm r}=\gamma^{\rm nr}+\gamma^{\rm qr}+\gamma^{\rm qnr}$), while the red dashed line marks the strong-to-weak threshold $g_{\rm th}$. Notably, whenever PA occurs, the efficiency increases significantly.
    (c) $\max(\eta_j)$ as a function of $\gamma^{\rm r}$ and $\gamma^{\rm qr}$, at fixed $g_{\rm eff}$, in the presence of both $\gamma^{\rm nr}$ and $\gamma^{\rm qnr}$, fixed at the values marked by the white dashed lines.
    The parameters are: (a,\,b) $\gamma_{1,2}^{\rm qr}/\tilde\omega_r=1.25\times10^{-4}$, $\gamma^{\rm nr}/\tilde\omega_r = 1 \times 10^{-4}$, $\gamma^{\rm qnr} = 0$, and (c) $g_{\rm eff}=1\times10^{-3}$, $\gamma^{\rm nr} =7.5 \times10^{-5}$, $\gamma^{\rm qnr}/\tilde\omega_r = 2 \times 10^{-4}$. For all panels $\abs{A_d}/\tilde\omega_{r}=2.5\times10^{-6}$.}
    \label{fig6}
\end{figure}

Having illustrated PA-based efficiency optimization for two representative loss configurations, we now step back for a broader view, specifically on how $\max(\eta_j)$ varies across the full parameter space of coupling strengths and dissipation channels, both radiative and non-radiative. 

Figure~\ref{fig6}(a,\,b) addresses this question by mapping the maximum efficiency, $\max(\eta_j)$, and the minimum reflection, $\min(\abs{S_{11}})$, as functions of $g_{\rm eff}$ and $\gamma^{\rm r}$. Each point is obtained by optimizing over $\omega_d$ and $\Delta_\omega$; the resulting optimum is found to occur with $\omega_d$ resonant with the relevant hybrid mode, at the corresponding optimal detuning.
For simplicity, in these plots, we consider a setting with non-zero resonator non-radiative loss $\gamma^{\rm nr}$ (indicated by the vertical cyan dashed lines in \figref{fig6}(a,\,b)) and zero qubit non-radiative losses ($\gamma^{\rm qnr}=0$).
A clear reduction of $\max(\eta_j)$ appears in the region where PA is not reached, whose boundary is marked by the cyan dashed lines (both vertical and curved), reflecting the tight link between near-perfect feeding ($|S_{11}|\approx0$) and efficient conversion established throughout this work. This relation, remarkably, holds regardless of the coupling regime: entering the weak nonlinear coupling regime (red dashed lines) does not by itself limit the maximum achievable efficiency, as long as $|S_{11}|\approx0$ is maintained, consistent with the results of \secref{weakeffcoupl}. On the contrary, in the parameter regions where PA cannot be achieved, the maximum efficiency is severely limited. This robustness is particularly valuable from an experimental standpoint, since strong effective coupling remains difficult to reach in typical circuit-QED architectures, having so far been demonstrated in only a few realizations \cite{TomonagaNatureComm2025,WangNatureComm2025}.

Beyond the coupling regime, the loss configuration itself matters. As noted in \secref{resPT}, fulfilling the PT-symmetry condition for $H_{\mathrm{PA}}$ (i.e., $\gamma^{\rm r}=\gamma^{\rm nr}+\gamma^{\rm qr}+\gamma^{\rm qnr}$, green vertical dashed line) is not the optimal choice for maximizing $\eta_j$. As established in \secref{Hrzsubsp}, exploiting instead the Hermitian subspaces recovered at finite detuning allows one to disregard the loss-balance condition and, more importantly, to reach simultaneously the highest conversion efficiencies achievable.
This corresponds to the brightest region of the map in \figref{fig6}(a), where the detuning drives the relevant hybrid mode toward a more \textit{atom-like} character, suppressing the resonator's effective non-radiative loss, as demonstrated in \appref{app_eff_PA}. The relation between $\max(\eta_j)$ and perfect feeding is further confirmed by the direct comparison of \figref{fig6}(a) and \figref{fig6}(b): the regions of reduced efficiency closely track the areas of significant residual reflection, $\abs{S_{11}}>0$.

Finally, \figref{fig6}(c) turns to a more realistic scenario in which both photonic (vertical white dashed line) and qubit (horizontal white dashed line) non-radiative losses are present simultaneously, mapping $\max(\eta_j)$ as a function of $\gamma^{\rm r}$ and $\gamma^{\rm qr}$ at fixed $g_{\rm eff}$, again optimized over $\omega_d$ and $\Delta_\omega$, corresponding to the relevant hybrid mode resonance. Although such losses would typically degrade the conversion efficiency, even in this case relaxing the strict PT-symmetry condition for $H_{\mathrm{PA}}$ (green line) and choosing a suitable choice of $\Delta_\omega$ significantly improve $\max(\eta_j)$ while keeping $\abs{S_{11}}\approx0$ (irrespective of the coupling regime) through a suitable choice of $\Delta_\omega$. Specifically, as previously illustrated, varying the detuning drives the relevant hybrid mode toward a more photon- or atom-like character, thus compensating the different subsystems' losses.
Here, since the qubits are assumed to carry the larger non-radiative loss rate, the global maximum efficiency is reached on the photon-like side ($|C_{j2}|^2<1/2$), away from the PT-symmetry condition, which corresponds to equal photonic and atomic fractions.

\section{Strong coupling between a single-photon and a two-photon Fock state}\label{SCfockstates}

Recently, SC between a two-photon and a one-photon state has been achieved in USC circuit QED \cite{WangNatureComm2025}, using a flux qubit embedded in a $\lambda/2$ coplanar waveguide resonator, where the qubit acts as a nonlinear mediator between two resonator modes. In this implementation, however, the radiative loss rates of the interacting modes cannot be independently engineered. To overcome this limitation, we propose an alternative setup in which two LC resonators are coupled to a common flux qubit acting as a nonlinear coupler, allowing the radiative losses of the individual modes and the strength of the nonlinear interaction to be tuned separately. In this section, we apply our non-Hermitian framework to this system, investigating near-deterministic down-conversion mediated purely by photon-photon interactions.

The coherent exchange between a single photon in mode $n=2$ and a photon pair in mode $n=1$ (with $\tilde\omega_2 \simeq 2\tilde\omega_1$), enabled by this system, is effectively described by the Hamiltonian \cite{KockumPRA2017, WangNatureComm2025}
\begin{equation}\label{H_dc}
    H_{\rm dc} = \tilde\omega_1 a_1^\dagger a_1 + \tilde\omega_2 a_2^\dagger a_2 + g_{\rm dc} \left[ a_1^2 a_2^\dagger + (a_1^\dagger)^2 a_2 \right]\,,
\end{equation}
where $a_1$ and $a_2$ are the annihilation operators for the fundamental and second-harmonic modes, with dressed frequencies $\tilde\omega_1$ and $\tilde\omega_2$, respectively, and $g_{\rm dc}$ is the effective coupling strength of this nonlinear process. Since $H_{\rm dc}$ commutes with the generalized excitation number operator $N_{\rm exc} = a_1^\dagger a_1 + 2a_2^\dagger a_2$ (the same concept as in \secref{onephtwatoms}, here weighted to reflect the two-photon nature of mode $2$), the Hilbert space block-diagonalizes accordingly. Denoting by $\ket{n_1,n_2}$ the photon population in modes $n=1$ and $n=2$, the vacuum $\ket{0,0}$ and the single-photon state $\ket{1,0}$ span the decoupled zero-excitation ($N_{\rm exc}=0$) and single-excitation ($N_{\rm exc}=1$) sectors, while the coherent nonlinear exchange between the second-harmonic and fundamental modes occurs entirely within the $N_{\rm exc}=2$ manifold, spanned by $\{\ket{2,0}, \ket{0,1}\}$. The corresponding hybridized eigenstates take the form
\begin{align}
    \ket{\psi_+}&= \cos{\biggl(\frac{\alpha}{2}\biggr)} \ket{2,0} + \sin{\biggl(\frac{\alpha}{2}\biggr)} \ket{0,1}\,,\\
    \ket{\psi_-} &= \cos{\biggl(\frac{\alpha}{2}\biggr)} \ket{0,1} - \sin{\biggl(\frac{\alpha}{2}\biggr)} \ket{2,0}\,, 
\end{align}
with $\tan{(\alpha)} = 2\sqrt{2}\,g_{\rm dc}/\Delta'_{\omega}$, where $\Delta'_\omega = \tilde\omega_2 - 2\tilde\omega_1$ is the detuning between the interacting modes. These hybridized modes are formally analogous to the eigenstates in \cref{eq4,eq5}.

We consider a weak coherent drive applied to the second-harmonic mode $a_2$, operating in the low-power limit, where the mean photon number in the absence of nonlinear coupling remains much smaller than one ($\langle a_2^\dagger a_2\rangle_{g_{\rm dc}=0} \ll 1$). Under this assumption, the generalized two-mode quantum Rabi Hamiltonian describing the system in the USC regime (see Refs.~\cite{KockumPRA2017, WangNatureComm2025}) reduces to the effective Hamiltonian $H_{\rm dc}$ of \eqref{H_dc}, and the reflection spectra of the pumped resonator can be evaluated with the same formalism of \secref{theory_s}. Here, we denote by $\gamma^{\rm r}_{2}$ and $\gamma^{\rm r}_{1}$ ($\gamma^{\rm nr}_{2}$ and $\gamma^{\rm nr}_{1}$) the radiative (non-radiative) decay rates of the second-harmonic and fundamental modes, respectively.

\begin{figure}[t]
\centering
\includegraphics[width=\linewidth]{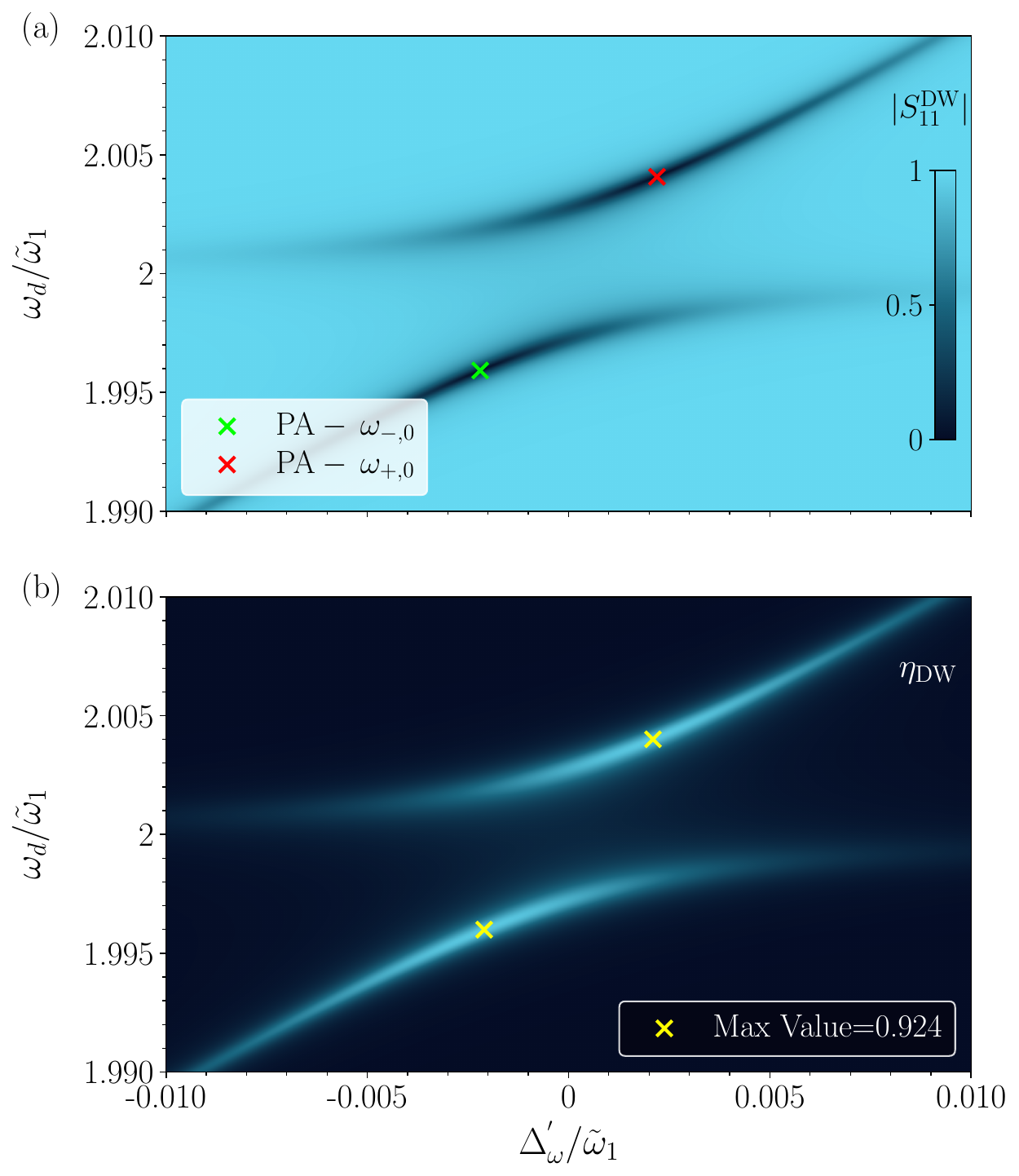}
    \caption{Single-photon to two-photon down-conversion in the photonic circuit-QED 
    system. 
    (a) Reflection coefficient $\abs{S^{\rm DW}_{11}}$ as a function of drive frequency $\omega_d$ and detuning $\Delta'_{\omega}$. Red and green crosses mark the PA points ($\abs{S^{\rm DW}_{11}}=0$) for the upper and lower hybrid-mode branches, respectively, each corresponding to the emergence of a Hermitian subspace of $H_{\rm PA}$. 
    (b) Spontaneous down-conversion efficiency $\eta_{\rm DW}$ as a function of $\omega_d$ and $\Delta'_{\omega}$. Yellow crosses indicate the maxima, $\eta_{\rm DW}\approx0.924$. 
    All results are obtained away from loss balance, $(\gamma^{\rm r}_2-\gamma^{\rm nr}_2) \neq 2(\gamma^{\rm r}_1+\gamma^{\rm nr}_1)$, with radiative losses dominating over the non-radiative ones and a strong non-radiative imbalance between the two modes, $\gamma^{\rm nr}_1\gg\gamma^{\rm nr}_2$. Specifically, we set $g_{\rm dc}/\tilde{\omega}_1 = 2\times10^{-3}$, 
    $\gamma^{\rm r}_1/\tilde{\omega}_1 = \gamma^{\rm r}_2/\tilde{\omega}_1 = 8\times10^{-4}$, 
    $\gamma^{\rm nr}_1/\tilde{\omega}_1 = 5\times10^{-5}$, 
    $\gamma^{\rm nr}_2/\tilde{\omega}_1 = 1.5\times10^{-5}$, 
    and $|A_d| = \gamma^{\rm r}_2/100$.}
\label{dcfigure}
\end{figure}

Figure~\ref{dcfigure}(a) shows the reflection spectra, computed as
\begin{equation}\label{s11dc}
    \abs{S^{\rm DW}_{11}(\omega_d)} = \abs{1 - \frac{\sqrt{\gamma^{\rm r}_{2}}}{A_d} \expec{a_2}_{\rm ss}}\,,
\end{equation}
where $\expec{a_2}_{\rm ss} = \Tr{a_{2}\rho_{\rm ss}}$ and $|A_d|^2$ is the input photon rate of mode $n=2$. The spectra are plotted as a function of the detuning $\Delta'_\omega$, tuned by varying the resonance frequency of the second-harmonic mode $\tilde\omega_2$; experimentally, this is achieved through the qubit flux offset, which shifts $\tilde\omega_1$ and $\tilde\omega_2$ with different slopes. 
The steady-state density matrix $\rho_{\rm ss}$ is obtained numerically as in \secref{theory_s}, by solving the master equation in \eqref{mastereq}, which retains the Lindblad structure of \eqref{liouv}, but with purely bosonic dissipators. Specifically, the atomic and resonator annihilation operators are replaced by those of the fundamental and second-harmonic modes, together with their loss rates $\gamma^{\rm r,nr}_{1,2}$. The resulting anticrossing between the hybrid modes (similar to that observed experimentally in Ref.~\cite{WangNatureComm2025}) confirms that the system operates in the QNLO SC regime for the chosen parameters.
The PA points, defined by $|S^{\rm DW}_{11}|=0$, are marked by red and green crosses for the upper and lower hybrid branches, respectively, and correspond, as discussed in \secref{Hrzsubsp}, to the emergence of a one-dimensional Hermitian subspace of $H_{\mathrm{PA}}$, defined analogously to \eqref{Hrz}. Here, the quantities $\bar{\Omega}_{j}$ entering $H_{\mathrm{PA}}$ are the eigenvalues of the matrix describing the closed-system transitions, i.e.,
\begin{equation}\label{Adc}
    A = \begin{pmatrix} 2 \tilde\omega_{1} & \sqrt{2}g_{\rm dc} \\ \sqrt{2}g_{\rm dc} & \tilde\omega_{2} \end{pmatrix}\,,
\end{equation}
while the effective loss terms read
\begin{equation}\label{gammadc}
    \tilde\gamma_j = 2(\gamma^{\rm r}_{1} + \gamma^{\rm nr}_{1})|C_{j1}|^2 + (\gamma^{\rm nr}_{2}-\gamma^{\rm r}_{2})|C_{j2}|^2\,,
\end{equation}
with $j=\{1,2\}$ labeling the upper and lower hybrid modes. As in \secref{Hrzsubsp}, the detuning tunes the hybrid-mode composition, encoded in the weights $|C_{j1}|^2$ and $|C_{j2}|^2$, until $\tilde\gamma_j$ vanishes at the marked crosses, where the corresponding eigenvalue of $H_{\mathrm{PA}}$ becomes purely real, and the reflection vanishes once the drive frequency is tuned to match it, i.e., $\omega_d=\bar{\Omega}_j$. If, in addition, the decay rates satisfy $(\gamma^{\rm r}_{2} - \gamma^{\rm nr}_{2}) = 2(\gamma^{\rm r}_{1} + \gamma^{\rm nr}_{1})$, this vanishing condition is met at zero detuning ($\Delta'_{\omega}=0$) for both hybrid modes simultaneously, and $H_{\mathrm{PA}}$ realizes a PT-symmetric configuration, with all the eigenvalues real (not shown here).

Figure~\ref{dcfigure}(b) shows the conversion efficiency map, defined as
\begin{equation}\label{effdc}
    \eta_{\rm DW} = \frac{\Phi_{\rm out}^{(1)}}{2 \Phi_{\rm in}^{(2)}} = \frac{\gamma^{\rm r}_{1} \expec{ a_1^\dagger a_1}_{\rm ss}}{2 |A_d|^2}\,,
\end{equation}
where $\Phi_{\rm out}^{(1)}= \gamma^{\rm r}_{1} \expec{ a_1^\dagger a_1}_{\rm ss}$ is the output photon rate of the fundamental mode and $\Phi_{\rm in}^{(2)} = |A_d|^2$ is the input pump photon rate of the second harmonic. The factor of $2$ accounts for the two-photon nature of the process, in which a single pump photon is down-converted into a fundamental photon pair, so that $\eta_{\rm DW}=1$ corresponds to perfect conversion. Exploiting PA together with the general framework established throughout this work, and the independent tunability of the two modes' losses discussed above, the spontaneous down-conversion efficiency at the single-photon level can be greatly enhanced. For the set of parameters chosen, for instance, the efficiency reaches about $92\%$ (yellow crosses), despite non-negligible non-radiative losses. These results confirm that the PA-based optimization strategy developed for the two-qubit platform, as well as all the previously discussed properties, extends directly to a purely photonic implementation.

\section{Discussions and Outlook}

In this work, we have shown that non-Hermitian physics offers a powerful and broadly applicable framework for accessing near-deterministic quantum nonlinear regimes. We first applied this approach to setup (i), a driven-dissipative circuit-QED system enabling the simultaneous excitation of two atoms by a single photon. We demonstrated that PA, i.e., the condition of zero reflection, acts as a natural spectral mechanism for optimizing single-photon nonlinear processes. Under strict loss-balance conditions, PA occurs at zero detuning for both hybrid modes generated by the nonlinear interaction, reflecting the exact PT symmetry of the effective non-Hermitian Hamiltonian $H_{\mathrm{PA}}$. In this ideal limit of the absence of non-radiative dissipation, the conversion efficiency reaches unity, the two-qubit output correlation function exhibits its maximal quantum value, and the cross-correlation spectra reveal maximally non-classical, pair-correlated photon emission.

When the strict loss-balance condition is relaxed, PA can be restored by introducing a finite detuning between the system's components. This effect originates from the emergence of a Hermitian subspace within $H_{\mathrm{PA}}$, enabled by reshaping the hybrid-mode composition through the Hopfield coefficients, which suppresses the impact of atomic or photonic non-radiative losses and substantially enhances the emission efficiency. Notably, we have demonstrated that the efficiency maximum coincides exactly with PA whenever the resonator is free of non-radiative loss, regardless of how large or asymmetric the qubit losses are. A phase-diagram analysis (in \secref{phasediagrams}) further shows that this strategy is especially effective when radiative losses exceed the corresponding non-radiative ones, and when intrinsic dissipation is predominant in one subsystem. These efficiency gains are mirrored by enhanced two-qubit correlations and pair-correlated emission spectra, consistent with the ideal PT-symmetric limit, and extend straightforwardly to non-identical qubits. Moreover, they persist even in the weak-nonlinearity regime ($g_{\rm eff} < g_{\rm th}$), provided $g_{\rm eff}$ remains above the minimum threshold $g_{\rm min}$ required for PA. This confirms that non-Hermitian engineering is not restricted to the SC domain or PT-symmetric settings, but is directly relevant for current and near-term circuit-QED implementations.

We then applied the same framework to setup (ii), where SC between a single-photon and a two-photon Fock state can be engineered, as recently demonstrated in circuit QED~\cite{WangNatureComm2025}. By exploiting PA and the Hermitian-subspace condition, spontaneous down-conversion at the single-photon level can reach near-unity efficiency even in the presence of appreciable non-radiative losses. This approach can be naturally extended to optical-frequency platforms, including nonlinear microresonators~\cite{Guo2016PRL,WangPRAWhispGall2023, ZhiYan2025} and integrated nanophotonic circuits~\cite{ZhaoPRL2020,McKennaNatureComm2022}, which exhibit strong $\chi^{(2)}$ nonlinearities accessible under continuous-wave driving. It can also be applied to three-wave-mixing scenarios, where the effective single-photon nonlinear coupling between two modes can be significantly enhanced by coherently driving a third interacting mode into a macroscopic coherent state~\cite{IrvinePRL2006}.

These results establish non-Hermitian engineering as a versatile and experimentally accessible strategy for optimizing single-photon nonlinear processes across a wide range of platforms and parameter settings.
While our analysis has focused on the frequency domain, under continuous-wave driving, extending this PA-based optimization to pulsed protocols is a natural next step, potentially enabling deterministic, single-shot state transfer directly in the time domain. By linking PA to the emergence of Hermitian subspaces and to an effective suppression of losses, our framework provides a systematic route to near-deterministic quantum frequency conversion, with potential implications for the generation of non-classical light, entangled photon pairs, quantum gates, and scalable quantum information processing at the single-photon level.

\begin{figure*}[!h]
    \centering
    \includegraphics[width=0.965\textwidth]{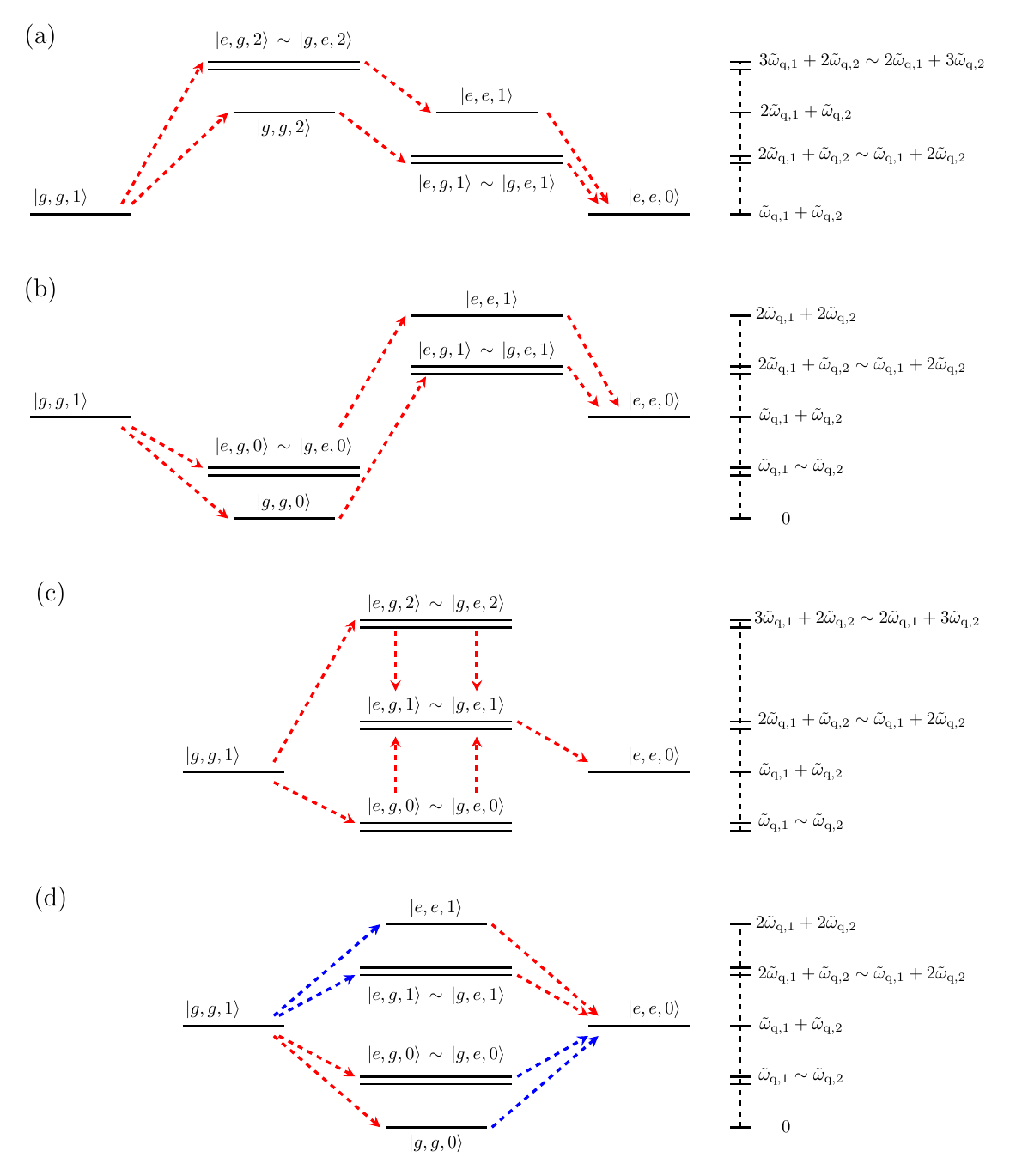}
    \caption{Virtual transition pathways between the states $|g,g,1\rangle$ and $|e,e,0\rangle$, contributing to the effective coupling $g_{\rm eff}$. Red arrows denote processes originating from counter-rotating terms of the light-matter interaction (first term in \eqref{hint}), while blue arrows indicate additional pathways induced by the spin-spin term (last contribution in \eqref{hint}). The corresponding energy-level diagram is shown on the right. For clarity, all energies are referred to the ground state $\ket{g,g,0}$, taken as the zero of energy. The two dressed qubit transition frequencies are considered non-identical ($\tilde{\omega}_{{\rm q},1}\neq\tilde{\omega}_{{\rm q},2}$), hence states such as $|e,g,n\rangle$ and $|g,e,n\rangle$ are non-degenerate. For compactness, the symbol $\sim$ groups them into a single entry (e.g.\ $|e,g,0\rangle\sim|g,e,0\rangle$): the associated transition path applies to both of these states, with the corresponding energy of each one labeled by the same notation on the right (e.g.\ $\tilde{\omega}_{{\rm q},1}\sim\tilde{\omega}_{{\rm q},2}$).}
    \label{fig8}
\end{figure*}

\appendix
\begin{figure}[b]
    \centering
    \includegraphics[width=\linewidth]{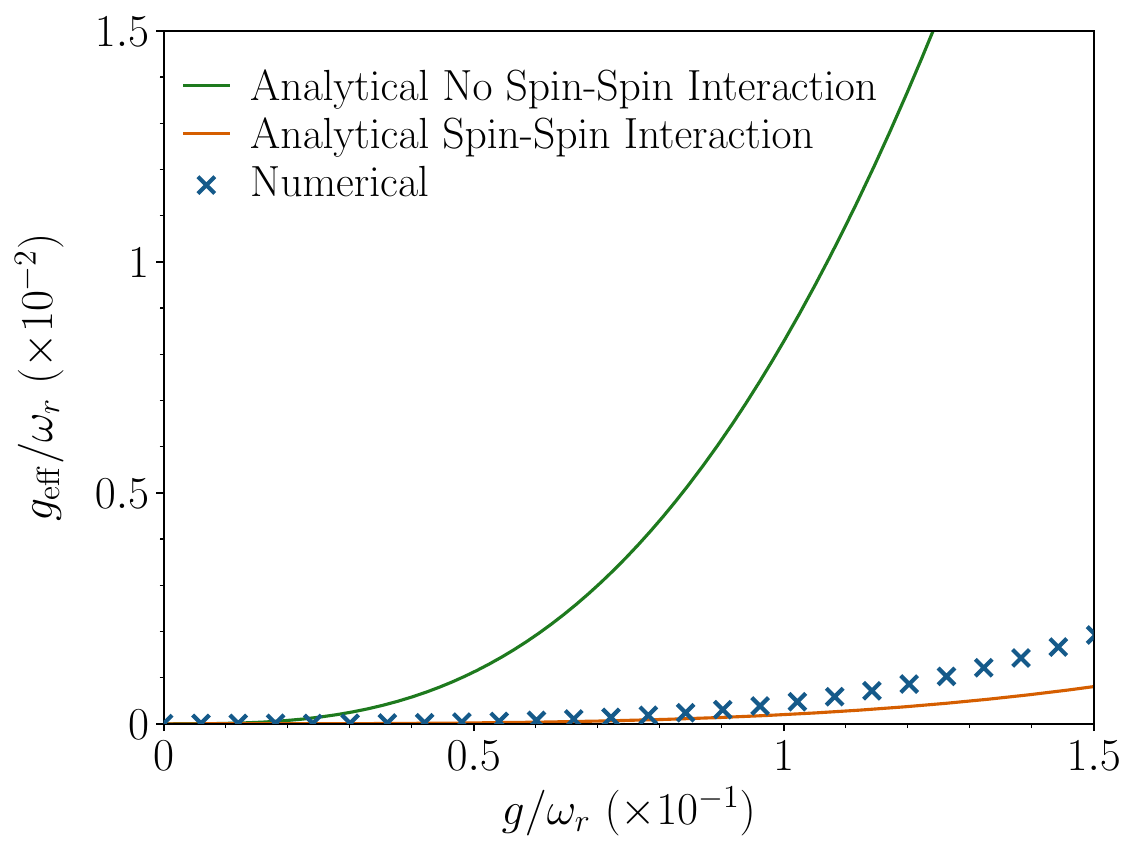}
    \caption{Effective coupling strength estimations with and without spin-spin 
interaction. 
Effective coupling $g_{\rm eff}$ as a function of bare coupling strength $g$ (with 
$g_1\equiv g$ and $g_2\equiv1.04\,g$), both normalized by the resonator frequency 
$\omega_r$. The numerical $g_{\rm eff}$ (blue crosses) is extracted as half the splitting at the  
avoided-crossing between the hybrid modes originating from 
$\ket{g,g,1}$ and $\ket{e,e,0}$, evaluated near resonance 
($\tilde{\omega}_r\simeq\tilde{\omega}_{{\rm q},1}+\tilde{\omega}_{{\rm q},2}$). 
It is compared with the analytical model from standard perturbation theory 
(see \eqref{geff_n_order}), including (orange curve) and excluding (green 
curve) the virtual transitions induced by the spin-spin interaction (last term 
in \eqref{hint}).}
    \label{fig9}
\end{figure}
\section{Effective Coupling in the Dicke Model with Spin-Spin interaction term}\label{g_eff_app}
In this appendix, we investigate how the spin-spin interaction introduced in the generalized Dicke Hamiltonian [see \eqref{Hams}] modifies the effective coupling ($g_{\rm eff}$) between the states $\ket{e,e,0}$ and $\ket{g,g,1}$. Since we focus on the near-resonant region, $\tilde{\omega}_r \simeq \tilde{\omega}_{{  \rm q},1} + \tilde{\omega}_{{\rm q},2}$, we employ standard degenerate perturbation theory, from which $g_{\rm eff}$ is obtained as
\begin{equation}\label{geff_n_order}
g_{\mathrm{eff}} = \sum_{j_1, j_2, \dots, j_{n-1}} \frac{V_{f j_{n-1}} \dots V_{j_2 j_1} V_{j_1 i}}{(E_i - E_{j_1})(E_i - E_{j_2}) \dots (E_i - E_{j_{n-1}})}\,,
\end{equation}
where the sum runs over all virtual transition pathways of any order $n$ connecting $\ket{i} = \ket{g,g,1}$ to $\ket{f} = \ket{e,e,0}$, through $n-1$ intermediate states $\ket{j_1}, \ket{j_2}, \dots, \ket{j_{n-1}}$. In \eqref{geff_n_order}, $E_k$ denotes the energy of the state $\ket{k}$, and $V_{km} = \bra{k} {H}_{\mathrm{int}} \ket{m}$. A complete derivation of this perturbative expression can be found in Ref.~\cite{KockumPRA2017}.
For our system, 
\begin{equation}\label{hint}
{H}_{\mathrm{int}}=- (g_1\Lambda_1 - g_2\Lambda_2)\, (a + a^\dagger) - 2\, \frac{g_1 g_2}{\omega_{r}}\, \Lambda_1 \Lambda_2\,,
\end{equation}
where $\Lambda_j = \cos{\theta_j}\, \sigma^x_j + \sin{\theta_j}\, \sigma^z_j$ (see \secref{onephtwatoms} for details). We compute $g_{\rm eff}$ using \eqref{geff_n_order}, where the virtual paths and the associated energy diagram are illustrated in \figref{fig8}. In particular, the pathways contributing to $g_{\rm eff}$ are of two types. Panels (a-c) show three-step paths, while panel (d) displays two-step ones. The red arrows denote virtual transitions driven by the counter-rotating terms of the light-matter interaction [first term in \eqref{hint}], which are linear in the coupling strengths $g_1$ and $g_2$. Hence, the virtual pathways in (a-c) each yield an overall contribution of order $g^{3}$. In panel (d), conversely, each path combines one red and one blue arrow, with the blue one induced by the spin-spin interaction [last term in \eqref{hint}], which scales as $g^{2}$. Together with the red arrow ($\sim g$), each of these mixed pathways also yields an overall contribution of order $g^{3}$. All processes in \figref{fig8} therefore contribute to $g_{\rm eff}$ at the same order ($g^3$).

In \figref{fig9}, we compare the numerically calculated $g_{\rm eff}$ with the analytical predictions as functions of the normalized bare coupling $g/\omega_{r}$, where $g_2 = 1.04\,g_1$ (with $g \equiv g_1$). The numerical values are extracted as half the splitting at the avoided level crossing between the hybridized states $\ket{\psi_{\pm }}$ in the spectrum of $H_{\rm S}$ (see \figref{fig_avoided_crossing}), while the analytical results are obtained from \eqref{geff_n_order}. Specifically, the model including spin–spin interaction paths (orange curve) agrees more closely with the numerical data (blue crosses) than the model without these virtual transitions (green curve), which consistently overestimates $g_{\rm eff}$. For the findings in \figref{fig9}, we set the parameters as in Ref.\,\cite{TomonagaNatureComm2025}. For qubit 1, $\tilde\omega_{{\rm q},1} = \sqrt{\epsilon_1^2 + \Delta_1^2}$, where $\epsilon_1$ is systematically chosen at the point of minimal gap in the avoided level crossing of the hybrid modes, and $\Delta_1/2\pi = 1.32$ GHz. For qubit 2, $\tilde\omega_{{\rm q},2} = \sqrt{\epsilon_2^2 + \Delta_2^2}$, with $\Delta_2/2\pi = 1.27$ GHz and $\epsilon_2/2\pi = -3.22$ GHz. Finally, we set $\theta_j = -\arctan(\epsilon_j / \Delta_j)$ and $\tilde \omega_{\rm r} = \tilde\omega_{{\rm q},1} + \tilde\omega_{{\rm q},2}$.

It would be of great interest to extend this analysis into a comprehensive study, exploring how opposing signs in light-matter coupling strengths and flux offsets can be leveraged for quantum technologies. Such configurations could be used to selectively suppress or enhance the effective strength of nonlinear interactions.

\section{Reflection spectrum formula and Non-Hermitian Hamiltonians}\label{appA}
In this appendix, we derive the reflection spectrum ($\abs{S_{11}}$) formula presented in the main text [see \eqref{s11}]. To this end, we adopt the QLE approach in the low-excitation regime. We also obtain the non-Hermitian Hamiltonian $H_{\mathrm{PA}}$ and highlight its connection to the zeros of $\abs{S_{11}}$. 

We begin with the total effective Hamiltonian $H_{\rm tot}$ describing the system in \figref{fig_modello_fin}:
\begin{equation}\label{totalH}
H_{\rm tot} = H_{\rm eff} + H_{B} + H_{I, {\rm eff}}\,,
\end{equation}
where $H_{\rm eff}$ is the dressed Hamiltonian of the two-qubit system and the resonator, including the coherent effective coupling between a single photon and two qubits [see \eqref{Heff}]. In \eqref{totalH}, the Hamiltonian of the reservoirs (or baths) is defined as
\begin{equation}
{H}_B = \int_{-\infty}^{+\infty}\! \!\!\!\!\! \!\!\!d\omega \, \omega \sum_{\rm k = r,\rm nr}\left({c}^\dagger_{\rm k}(\omega) {c}_{\rm k}(\omega)+ \sum_{j=1}^2 {d}^\dagger_{j,{\rm k}}(\omega) {d}_{j,{\rm k}}(\omega) \right)\,,
\end{equation}
 where ${c}_{\rm k}(\omega)$ and ${d}_{j,{\rm k}}(\omega)$ denote the bosonic annihilation operators of the baths coupled to the resonator and to the $j$-th qubit, respectively. Specifically, the index $\rm k= r$ refers to the baths representing the input-output transmission lines (see \figref{fig_modello_fin}), while $\rm k= nr$ is associated with the reservoirs corresponding to non-radiative losses. Under the rotating-wave approximation, the effective system-bath interaction Hamiltonian reads
\begin{equation}
\begin{split}
{H}_{I,{\rm eff}} &= i\int_{-\infty}^{+\infty} \!\!\!
d\omega \,\left[\sum_{\rm k=r,{\rm nr}} \sqrt{\frac{\gamma^{ \rm k}}{2\pi}}( c_{\rm k}(\omega)  a^\dagger- c^\dagger_{\rm k}(\omega)\,  a)\right]\\
&+\left[\sum_{{\rm k} =\mathrm{r, nr}} \sum_{j=1}^2\sqrt{\frac{\gamma^{\rm q{\rm k}}_{j}}{2\pi}}\, ( {d}_{j,{\rm k}}(\omega)  \sigma^+_j- {d}^\dagger_{j,{\rm k}}(\omega) \sigma^-_j)\right]\,,
\end{split}
\end{equation}
where $\gamma^{\rm k}$ and $\gamma^{\rm q{\rm k}}_{j}$ are the loss rates of the resonator mode and the $j$-th qubit, respectively (see \secref{theory_s} and \tabref{table1} for more details). Note that these rates are responsible exclusively for the decay transitions within the subspaces shown in \figref{fig_avoided_crossing}(b,\,c), and do not represent the bare losses of the entire light or matter subsystem. We derive the QLEs by substituting the bath operator solutions into the Heisenberg equations of motion for the system \cite{gardiner2004quantum, walls2008quantum}. For the resonator annihilation operator, we have
\begin{equation}\label{langres}
\begin{split}
\frac{d}{dt}{a} = &\left[- i \tilde\omega_{{r}} - \left(\frac{\gamma^{\mathrm{r}}+ \gamma^{\rm nr}}{2}\right)\right] {a} - i g_{\mathrm{eff}} {\sigma^-_1 \sigma^-_2} \\
&+ \sum_{\rm k = r,\rm nr}\sqrt{\gamma^{\mathrm{k}}} \, {a^{\rm (k)}_{\rm in}}\,,
\end{split}
\end{equation}
while for the two-qubit lowering operator we obtain
\begin{equation}\label{langqu}
\begin{split}
 \frac{d}{dt}{({\sigma^-_1 \sigma^-_2})}  &= 
\left[ -i 2 \tilde\omega_q - \frac{(\gamma^{\mathrm{qr}} +\gamma^{\mathrm{qnr}})}{2} \right] {\sigma_1^- \sigma_2^-} \\
&+i g_{\mathrm{eff}} {a\, (\sigma^+_1 \sigma^-_1 \sigma^{z}_2 + \sigma^{z}_1 \sigma^-_2 \sigma^+_2)}\\
&+\sum_{k =\mathrm{r, nr}} \sum_{j=1}^2 \sqrt{\gamma_j^{\rm qk}} \, \sigma^{z}_{j} \sigma_{m}^{-} \, b^{\,({\rm k})}_{j,{\rm in}}\,, \quad (m \neq j)\,.
\end{split}
\end{equation}
The input fields are written as 
\begin{equation}\label{input-field}
\begin{split}
&a^{(\rm k )}_{\rm in}(t)= \frac{1}{\sqrt{2 \pi}}\int_{-\infty}^{+ \infty} d\omega \,e^{-i \omega \, (t-t_0)}  c_{\rm k}(\omega; t_0)\;,\\
& b^{\,({\rm k})}_{j,{\rm in}}(t)=\frac{1}{\sqrt{2 \pi}}\int_{-\infty}^{+ \infty} d\omega \,e^{-i \omega \, (t-t_0)}  d_{j, {\rm k}}(\omega;t_0)\;,
\end{split}
\end{equation}  
where the index $\rm k$ indicates the radiative $(\rm k = r)$ and the non-radiative $(\rm k = nr)$ contributions. Moreover, $t_0$ represents an initial reference time ($t_0 < t$) at which the input reservoir operators are evaluated to solve the system dynamics. Similarly, the output fields are expressed by analogous expressions, with the final time $t_f$ ($t < t_f$) taking the place of $t_0$ to account for the field evolution long after the system-bath interaction. The QLEs can also be formulated in terms of the output operators \cite{walls2008quantum, gardiner2004quantum}, rather than the input ones shown in \eqref{langres} and \eqref{langqu}. By combining the input- and output-based forms of the QLEs in the frequency domain (referenced to the drive frequency $\omega_d$), we obtain the input-output relations in \eqref{inoutres} and \eqref{inoutqs}. The difference in sign between the last terms of these two equations originates from the relation $\sigma_j^z \sigma_j^- = -\sigma_j^-$. 

In our scenario (see \secref{theory_s} and \figref{fig_modello_fin}), a weak coherent drive is applied to the input channel of the resonator. Consequently, $\langle a^{\mathrm{(nr)}}_{\mathrm{in}} \rangle = \langle \, \sigma^{z}_{j} \, \sigma_{m}^{-} \,b^{\,(\mathrm{r)}}_{j,\mathrm{in}} \rangle = \langle \sigma^{z}_{j} \,\sigma_{m}^{-}\,b^{\,(\mathrm{nr)}}_{j,\mathrm{in}} \rangle = 0$ (with $j \in \{{1,2}\}$ and $m \neq j$), while $\lvert \langle a^{\mathrm{(r)}}_{\mathrm{in}} \rangle \rvert = \lvert A_d \rvert$. Under these conditions, the resonator reflection coefficient is given by
$$\abs{S_{11}(\omega_d)} = \abs{\frac{\expec{a^{(\rm r )}_{\rm out}(\omega_d)}}{{\expec{a^{(\rm r )}_{\rm in}(\omega_d)}}}}\,.$$ We note that the term proportional to $i g_{\rm eff}$ in \eqref{langqu} can be rewritten as $a(\ket{e,e} \bra{e,e}-\ket{g,g}\bra{g,g})$. In the low-excitation regime, we approximate its expectation value by $\langle -a \rangle$. Linearizing the QLEs in the low-excitation limit and Fourier-transforming to the frequency domain, we obtain
\begin{equation}\label{langeqm}
-i\omega_d\,{{\beta}}(\omega_d) = -i A\,{{\beta}}(\omega_d) - \frac{1}{2}{\Gamma_{\rm pol}}\,{{\beta}}(\omega_d)+ {{F}}_{\rm in}(\omega_d)\,,
\end{equation}
where
\begin{equation}\label{matrix_multispin}
 {\beta}=\begin{pmatrix}
\expec{{a}(\omega_d)}&\\
 \expec{(\sigma^-_1 \circledast \,\sigma_2^-)(\omega_d)}&\\
\end{pmatrix} \,,\,{A} = \begin{pmatrix}
\tilde \omega_r & {g_{\rm eff}} \\
{g_{\rm eff}} & {2\tilde\omega_{\rm q}}
\end{pmatrix} \, ,\; 
\end{equation} 
\begin{equation}
{\Gamma_{\rm pol}} =
\begin{pmatrix}
 \gamma^{\rm r}  +  \gamma^{\rm nr} & 0 \\
0 &   \gamma^{\rm qr} + \gamma^{\rm qnr}
\end{pmatrix}
\,,\,{F_{\rm in}} = \begin{pmatrix}
\sqrt{ \gamma^{\rm r}}\expec{a^{\rm (r)}_{\rm in}(\omega_d)}\,   \\
0 
\end{pmatrix} \,.
\end{equation}
The symbol $\circledast$ in \eqref{matrix_multispin} denotes  convolution in the frequency domain.

In the low-excitation regime, the nonlinear QLEs [\eqref{langres} and \eqref{langqu}] can be mapped onto a linear bosonic-like system. Under these assumptions, an analytical expression for $\abs{S_{11}}$ can be derived. By solving \eqref{matrix_multispin} for $\langle a(\omega_d) \rangle$ in terms of $\langle a_{\rm in}^{(r)}(\omega_d) \rangle$, and substituting the result into the input--output relation in \eqref{inoutres}, after straightforward algebra one recovers the reflection coefficient in \eqref{s11}.

As described in \secref{theory_s}, the complex frequencies $\Omega_{\pm}$ are the poles of $\abs{S_{11}}$ and coincide with the eigenvalues of $H_{\rm pol}$ [see \eqref{hres}], obtained by incorporating the loss terms $\Gamma_{\rm pol}$ into the system Hamiltonian. By calculating these eigenvalues, we find
\begin{equation}\label{polfreq}
\begin{split}
&\Omega_{\pm} = \frac{\tilde \omega_{r}}{2}+\tilde\omega_{\rm q}+\-\frac{i}{4}\biggl[ \gamma^{\rm r}+  \gamma^{\rm nr}+\sum^2_{j=1} (\gamma^{\rm qr}_{j} + \gamma^{\rm qnr}_{j}) \biggl]\\
&\pm\frac{1}{2}\sqrt{4\,g_{\rm{eff}}^2+\biggr\{\Delta_\omega-\frac{i}{2}\biggr[ \gamma^{\rm r}+ \gamma^{\rm nr}-\sum^2_{j=1}( \gamma^{\rm qr}_{j} +\gamma^{\rm qnr}_{j})\biggr] \biggr\}^2}\,.
\end{split}
\end{equation}
The dynamics of $\abs{S_{11}}$ zeros can be derived by requiring the PA in \eqref{langeqm}, i.e., $\expec{a^{\rm (r)}_{\rm out}}=0$, which implies $\expec{a^{(\rm r)}_{\rm in}} = \sqrt{ \gamma^{\rm r}}\expec{a}$ \cite{zhangNATCOMM2017}. In this condition, \eqref{langeqm} reduces to
\begin{equation}\label{a_med_pa}
-i\omega_d\,{{\beta}}(\omega_d) = -i {A}{{\beta}}(\omega_d) - \frac{1}{2}{\Gamma_{\rm PA}}\,{{\beta}}(\omega_d) \,.
\end{equation} 
The non-Hermitian Hamiltonian $H_{\mathrm{PA}}$, introduced in \eqref{hrz_bare} and expressed in the subspace spanned by the states $\{\ket{g,g,1}, \ket{e,e,0}\}$, leads to the equation of motion in \eqref{a_med_pa}. ${\Gamma_{\rm PA}}$ differs from $\Gamma_{\rm pol}$ by a sign reversal of $ \gamma^{\rm r}$, acting as an effective gain term in $H_{\mathrm{PA}}$ \cite{zhangNATCOMM2017}. The eigenvalues of $H_{\rm PA}$ coincide with the complex frequencies $\tilde \Omega_{\pm }$ [see \eqref{s11}]. As discussed in \secref{theory_s}, when $H_{\mathrm{PA}}$ is in the PT-unbroken phase, its spectrum is real and PA occurs for all eigenmodes at zero detuning $\Delta_\omega=0$. Beyond this special case, the 'Hermitian Subspaces' framework, employed in Ref.\,\cite{BonizzoniNatureComm2025}, provides a method to predict PA through the tunable composition of the hybrid modes, as quantified by the Hopfield coefficients. Following the same procedure for the linearized system in \eqref{a_med_pa}, we obtain $H_{\mathrm{PA}}$ in the hybrid-mode basis reported in \eqref{Hrz}. This representation is valid in the SC regime, specifically when $g_{\rm eff}$ is much larger than all loss rates (see \secref{theory_s} for details).

All results presented in the main text are obtained in the low-excitation regime, $\bar n(\omega_d)\ll1$, where $\bar n(\omega_d)$ denotes the mean resonator photon number for $g_{\rm eff}=0$. This quantity can be derived from \eqref{langeqm}, yielding
\begin{equation}\label{nph}
\bar n(\tilde \omega_d) = \expec{{a}^\dagger a}_{\rm ss}= \frac{4\, \gamma^{\rm r}\,\abs{A_d}^2}{4(\tilde\omega_{r}-\omega_d)^2+( \gamma^{\rm r}+  \gamma^{\rm nr})^2}\,.
\end{equation}

\section{Analytical derivation of emission efficiency and correlations}\label{appB}
In this appendix, we derive the master equation truncated to the subspace shown in Fig.~2(c) and use it to obtain an analytical expression for the steady-state density matrix $\rho_{\rm ss}$ of the system described in \secref{theory_s}. The resulting analytical model is then used to validate the emission properties discussed in \secref{resPT}. To derive a time-independent $\rho_{\rm ss}$, we move into the rotating frame of the drive via the following unitary transformation:
\begin{equation}
\begin{split}
H'_{ \rm eff} =&\; U^\dagger (H_{\rm eff} + H_d(t))U + i \dfrac{dU^\dagger}{dt}U \\
=& \sum_{j=1}^2 \frac{1}{2}\biggl(\tilde\omega_{{\rm q},\,j }-\frac{\omega_d}{2} \biggr)\,\sigma^{z}_j + (\tilde \omega_{r}- \omega_d )\, a^\dagger a \\
&+ g_{{\rm eff}}(a \,\sigma^+_1 \sigma^+_2+{\rm H.c.}) +\sqrt{ \gamma^{\rm r}} (- i \,A_d\, a+ i \, A^*_d \,a^\dagger)\,,
\end{split}
\end{equation}
where $U =\exp{-i\omega_dt(\frac{1}{4}\sigma^{z}_1+ \frac{1}{4}\sigma^{z}_2+ a^\dagger a)}$. The equation of motion for the density matrix is given in \eqref{mastereq}, where \begin{equation}\label{liouv}
\mathcal{L}_s \,\rho = (\gamma^{\rm r} + \gamma^{\rm nr})\mathcal{D}[a]\rho + \sum^2_{j=1}(\gamma^{\rm qr}_{j} + \gamma^{\rm qnr}_{j} )\mathcal{D}[\sigma^-_j]\rho\,,
\end{equation}
with $\mathcal{D}[O]\rho = \tfrac{1}{2} (2 \,O\rho O^\dagger-\rho O^\dagger O-O^\dagger O \rho)$ representing the Lindblad dissipator for a general system operator $O$ \cite{breuer2002theory, gardiner2004quantum,walls2008quantum}.

To derive a simple analytical form for the steady-state expectation value of the single-qubit population $\langle \sigma_j^+ \sigma_j^- \rangle_{\rm ss}$, we adopt a local truncation approach. In the following, $\rho_{mn} \equiv \langle m | \rho | n \rangle$ denotes a matrix element of the density operator $\rho$ in the relevant basis of system eigenstates [see \secref{onephtwatoms} and \figref{fig_avoided_crossing}(c)], with $m,n \in \{ 0, 1, -, + \}$, where $+$ and $-$ label the hybrid modes $\ket{\psi_\pm}$.
When the driving frequency $\omega_d$ is nearly resonant with a single hybrid mode branch (e.g., the lower one, ${\omega}_-$), the detuning from the other one (${\omega}_+$) is large in the SC regime. Under these assumptions, the off-diagonal coherence $\rho_{-+}$ becomes negligible, i.e., $\rho_{-+} \approx 0$, thereby suppressing interference terms between the two hybrid modes. Consequently, we truncate the Hilbert space to a three-level effective subsystem: the ground state ($|0\rangle$), the resonant hybrid mode (e.g., $|\psi_-\rangle$), and the intermediate symmetric state $|1\rangle \approx (|e,g,0\rangle 
+ |g,e,0\rangle)/{\sqrt{2}}$, populated via spontaneous emission from $\ket{\psi_-}$ (or $\ket{\psi_+}$). More precisely, under local qubit losses, the full five-level dissipative 
manifold, including the antisymmetric state, would be required to 
rigorously conserve probability. Here, however, we impose a restricted normalization condition $\rho_{00} + \rho_{11} + \rho_{--} = 1$ on 
the truncated subspace, where $\rho_{--}$ is replaced by $\rho_{++}$ when the upper hybrid mode is involved. 
Under these approximations and using the equation of motion from \eqref{mastereq}, we derive the following relations for the matrix elements of $\rho$:
\begin{equation}\label{blocheqs}
\begin{split}
\frac{d}{dt}\rho_{11} \,\,\,\approx& \, (\gamma^{\rm qr} + \gamma^{\rm qnr})\biggr[ -\frac{1}{4}\rho_{11} + \frac{1}{2} \cos^2{\left(\frac{\alpha}{2}\right)}\rho_{--}\biggl]\,, \\
\frac{d}{dt}\rho_{--} \approx& -( \gamma^{\rm r}+ \gamma^{\rm nr})\sin^2{\left(\frac{\alpha}{2}\right)}\rho_{--}-\cos^2{\left(\frac{\alpha}{2}\right)}\\
&\quad\,(\gamma^{\rm qr} + \gamma^{\rm qnr})\rho_{--}-\sqrt{ \gamma^{\rm r}}A_d (\rho_{0-}+\rho^*_{0-})\,,\\
\frac{d}{dt}\rho_{0-}\, \approx& -i(\omega_{0,-}+\omega_{d})\rho_{0-}+ \sqrt{ \gamma^{\rm r}}A_d\,\sin
\left(\frac{\alpha}{2}\right)\\
&(-\rho_{00}+ \rho_{--})-\frac{ \gamma^{\rm r}+ \gamma^{\rm nr}}{2}\sin^2{\left(\frac{\alpha}{2}\right)}\rho_{0-}-\\
&\frac{(\gamma^{\rm qr} + \gamma^{\rm qnr})}{2} \,\cos
^2{\left(\frac{\alpha}{2}\right)}\rho_{0-}\,,
\end{split}
\end{equation}
where $\rho^*_{mn}$ indicates the complex conjugate of the corresponding matrix element; see \tabref{table1} for a definition of the main physical quantities. An analogous set of equations for the upper hybrid mode is derived by interchanging $\sin(\alpha/2)$ and $\cos(\alpha/2)$ throughout \eqref{blocheqs}.

At steady state, we can write the analytical conversion efficiency, defined in \eqref{eff}, as
\begin{equation}\label{spsmlower}
\eta^{\rm th}_{j}(\omega_d) = \gamma^{\rm qr}_{j}\frac{\langle \sigma_j^+ \sigma_j^- \rangle_{\rm ss}^{\rm th}}{\abs{A_d}^2} \approx  \frac{2\gamma^{\rm qr}_{j}}{\abs{A_d}^2} \,\rho_{--} \, \cos^2{ \biggl(\frac{\alpha}{2}\biggr)}\,,
\end{equation}
for the lower hybrid mode. When $\omega_d \approx \omega_{0,+}$, the relevant contribution instead reads
\begin{equation}
\eta^{\rm th}_{j}(\omega_d)  = \gamma^{\rm qr}_{j}\frac{\langle \sigma_j^+ \sigma_j^- \rangle_{\rm ss}^{\rm th}}{\abs{A_d}^2} \approx\frac{2\gamma^{\rm qr}_{j}}{\abs{A_d}^2} \rho_{++} \, \sin^2{ \biggl(\frac{\alpha}{2}\biggr)}\,.
\end{equation}
\begin{figure}[t]
    \centering
    \includegraphics[width=\linewidth]{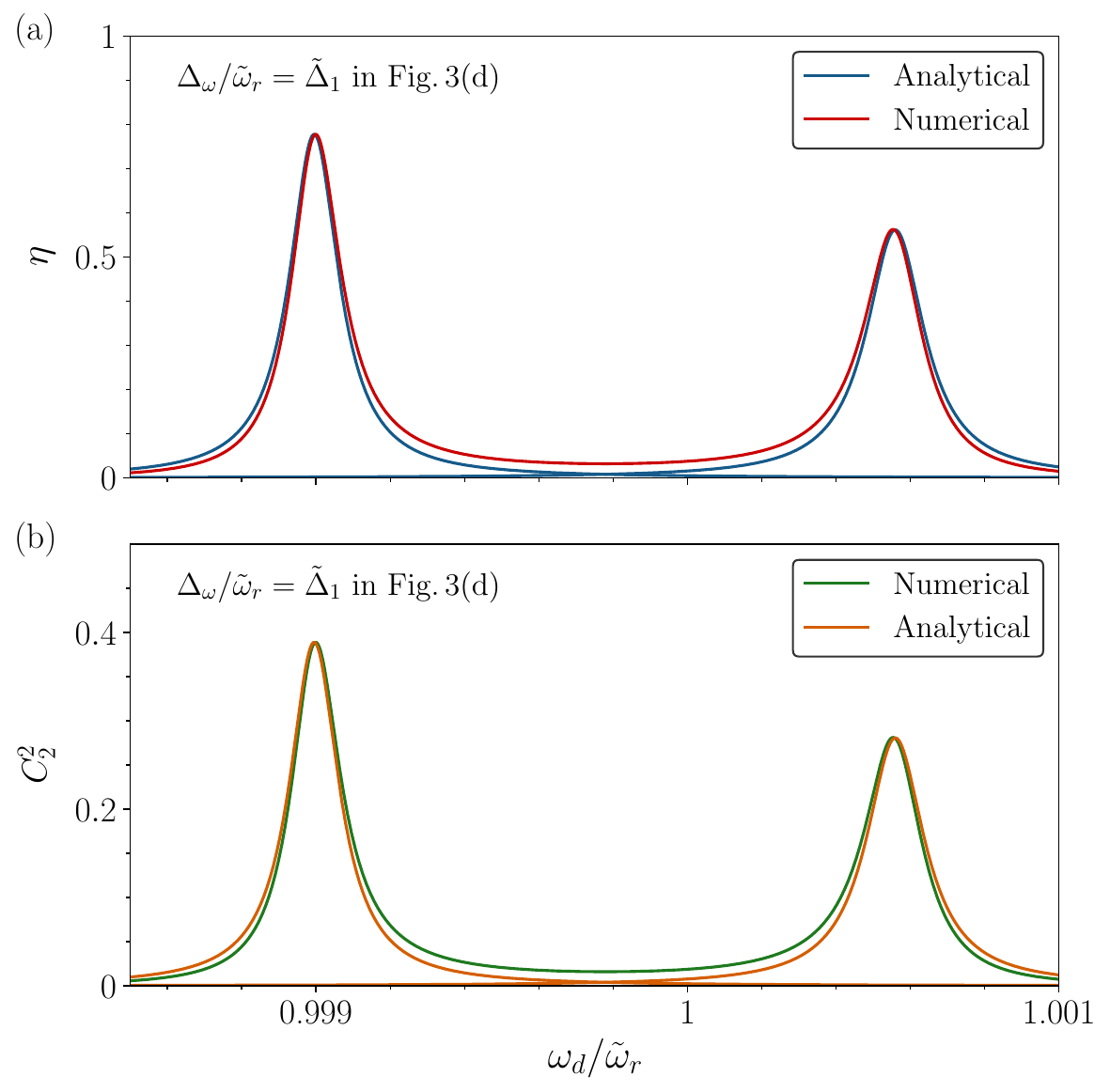}
    \caption{Analytical and numerical comparison of emission efficiency and two-qubit 
correlations for equal qubits. 
(a) Emission efficiency $\eta_j$ (with $\eta_1=\eta_2$) and (b) two-qubit output 
correlation function $C^q_2$ as functions of $\omega_d$. Whereas 
in \figref{fig1}(d) both $\omega_d$ and $\Delta_\omega$ are scanned, here 
$\Delta_\omega$ is fixed at one of the point of maximum efficiency, $\tilde\Delta_1$. The numerical results 
[red in~(a), green in~(b)] are compared with the analytical model [blue in~(a), 
orange in~(b)] obtained from the steady state of the master equation truncated 
to the subspace of \figref{fig_avoided_crossing}(c). The parameters used are the same as 
in \figref{fig1}(d).}
    \label{fig10}
\end{figure}
By imposing the steady-state condition, $d/dt\, (\,\rho_{mn})= 0$, on \eqref{blocheqs}, \eqref{spsmlower} can be explicitly expressed as
\begin{equation}
\eta^{\rm th}_{j}(\omega_d) \approx \frac{2 \gamma^{\rm qr
}_{j}\,  \gamma^{\rm r}\sin^2{(\frac{\alpha}{2})}\cos^2{(\frac{\alpha}{2})}}{(\omega_{d}-\omega_{-,0})^2+(\frac{1}{2}\gamma_2)^2}\,,
\end{equation}
where 
\begin{equation}
\gamma_2 =\biggl[( \gamma^{\rm r} + \gamma^{\rm nr})\,\sin^2{ \biggl(\frac{\alpha}{2}\biggr)}+ (\gamma^{\rm qr} + \gamma^{\rm qnr})\,\cos^2{ \biggl(\frac{\alpha}{2}\biggr)}\biggr]\,
\end{equation}
represents the lower hybrid-mode decay rate.
A similar Lorentzian expression for the upper mode is obtained by replacing the resonance $\omega_{-,0}$ and the decay rate $\gamma_2$ with $\omega_{+,0}$ and $\gamma_1$, the latter following from $\gamma_2$ by interchanging $\sin(\alpha/2)$ and $\cos(\alpha/2)$.
Following the same procedure, we find the analytical two-qubit output correlation function [see \eqref{corr}]:
\begin{equation}\label{c2qanal}
\begin{split}
 C^{q,{\rm th}}_{2}= \sqrt{\gamma^{\rm qr}_1 \gamma^{\rm qr}_2}\,\,  \gamma^{\rm r}\,
 \frac{\sin^2{(\frac{\alpha}{2})}\cos^2{(\frac{\alpha}{2})}}{(\omega_{d}-\omega_{-,0})^2+(\frac{1}{2}\gamma_2)^2}\,. 
 \end{split}
\end{equation}
In \figref{fig10}, we compare $\eta_j$ and $C_2^q$ obtained from \eqref{spsmlower} and \eqref{c2qanal} (and their upper-mode counterparts) with the corresponding numerical results, where $\Delta_\omega/\tilde\omega_r$ is fixed to $\tilde{\Delta}_1$ in \figref{fig1}(d). The excellent agreement between the numerical and analytical results confirms the validity of the discussion presented in \secref{resPT}.

\section{Relation between PA and maximum efficiency}
\label{app_eff_PA}

In this appendix, we show, starting directly from \eqref{gammabar} and \eqref{emisslow}, that the detuning maximizing $\eta_j$ coincides with the PA detuning (where $\tilde\gamma_j=0$) if and only if $\gamma^{\rm nr}=0$, irrespective of the qubit (either radiative or non-radiative) loss rates.

First, we recall the definitions of the effective loss rates entering $H_{\mathrm{PA}}$ in \eqref{gammabar} and the real hybrid-mode linewidths appearing in \eqref{emisslow}, namely
\begin{align}
    \tilde{\gamma}_j &= (- \gamma^{\rm r} +  \gamma^{\rm nr})|C_{j1}|^2 + (\gamma^{\rm qr} + \gamma^{\rm qnr})|C_{j2}|^2\,,\\
    \gamma_j &= ( \gamma^{\rm r} +  \gamma^{\rm nr})|C_{j1}|^2 + (\gamma^{\rm qr} + \gamma^{\rm qnr})|C_{j2}|^2\,,
\end{align}
related by the reversal of the sign of $\gamma^{\rm r}$. By writing $x \equiv |C_{j1}|^2$ for the photonic Hopfield weight of the $j$-th hybrid mode (so that $|C_{j2}|^2 = 1-x$), which can be varied through the detuning $\Delta_\omega$, these equations can be written equivalently as
\begin{align}
    \tilde\gamma_j(x) &= (\gamma^{\rm nr} - \gamma^{\rm r})\,x + (\gamma^{\rm qnr} + \gamma^{\rm qr})(1-x)\,,\\
    \gamma_j(x) &= (\gamma^{\rm nr} + \gamma^{\rm r})\,x + (\gamma^{\rm qnr} + \gamma^{\rm qr})(1-x)\,.
\end{align}

Perfect absorption of the $j$-th mode occurs at the value $x_{\rm PA}$ solving $\tilde\gamma_j(x_{\rm PA})=0$, i.e.,
\begin{equation}
x_{\rm PA} = \frac{\gamma^{\rm qr}+\gamma^{\rm qnr}}{\gamma^{\rm qr}+\gamma^{\rm qnr}+\gamma^{\rm r}-\gamma^{\rm nr}}\,.
\end{equation}

On the other hand, from \eqref{emisslow}, the efficiency reaches its maximum at the $j$-th hybrid-mode resonance (i.e., $\omega_d = \omega_{j,0}$), so that the maximum efficiency of the $k$-th qubit reduces to $\eta_k^{\rm max}(x) \propto x(1-x)/\gamma_j(x)^2$. Imposing $d\eta_k^{\rm max}/dx=0$, we obtain the condition (for both qubits)
\begin{equation}
x_{\eta} = \frac{\gamma^{\rm qr}+\gamma^{\rm qnr}}{\gamma^{\rm qr}+\gamma^{\rm qnr}+\gamma^{\rm r}+\gamma^{\rm nr}}\,.
\end{equation}
Comparing $x_{\rm PA}$ and $x_\eta$, the two photonic fractions (and thus detunings) coincide if and only if $\gamma^{\rm nr}=0$. Notably, this condition involves neither $\gamma^{\rm qr}$ nor $\gamma^{\rm qnr}$: whether the PA and efficiency-maximum detunings coincide depends entirely on the resonator's non-radiative loss, regardless of how large or asymmetric the two qubits' losses are. This accounts for the different behavior observed both in presence of PT symmetry in \figref{fig1}(a,\,b) (where $\gamma^{\rm nr}=0$) and \figref{fig1}(c,\,d) (where $\gamma^{\rm nr}\neq0$), as well as in the presence of Hermitian subspaces in \figref{fig4}(a,\,b) (where $\gamma^{\rm nr}\neq0$) and \figref{fig4}(c,\,d) (where $\gamma^{\rm nr}=0$).

\section{Non-equal qubit frequencies}\label{appD}

In this appendix, we examine how the spectral and emission properties 
discussed in \secref{theory_s} are modified when the two dressed qubit 
transition frequencies are nondegenerate, i.e., 
$\tilde{\omega}_{{\rm q},1}\neq\tilde{\omega}_{{\rm q},2}$.

\begin{figure}[t]
    \centering
    \includegraphics[width=\linewidth]{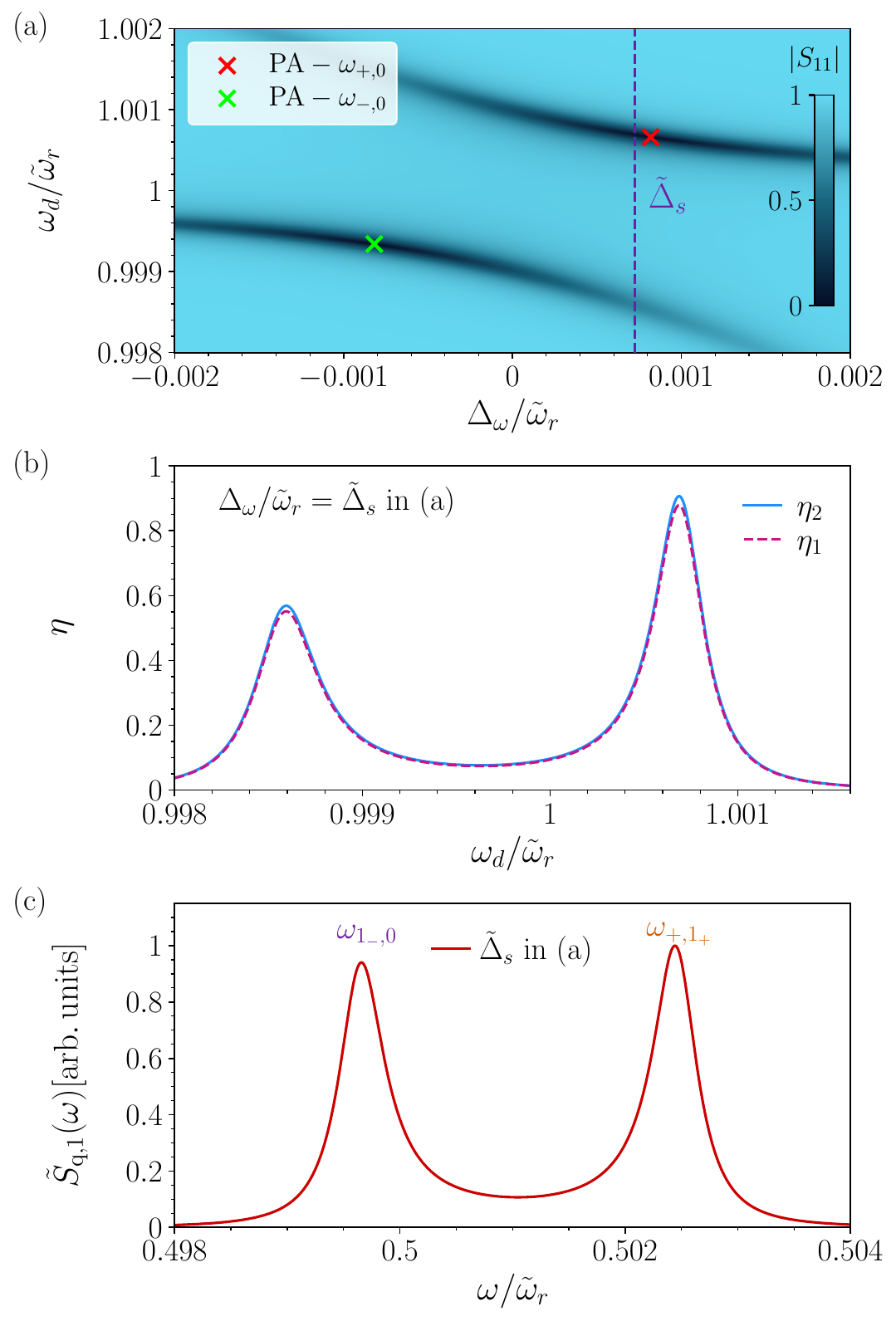}
    \caption{Spectral and emission properties for non-equal qubit frequencies 
($\tilde{\omega}_{{\rm q},1}\neq\tilde{\omega}_{{\rm q},2}$). All decay channels 
are included, with broken loss balance $\gamma^{\rm r}\neq\gamma^{\rm nr}+\gamma^{\rm qr}+\gamma^{\rm qnr}$. Radiative 
losses dominate over the non-radiative ones, and the qubit non-radiative decay rates 
are asymmetric ($\gamma^{\rm qnr}_1\neq\gamma^{\rm qnr}_2$) and exceed the 
resonator non-radiative one. 
(a) Reflection coefficient $\abs{S_{11}}$ as a function of drive frequency $\omega_d$ and detuning 
$\Delta_\omega = \tilde{\omega}_r - 
(\tilde{\omega}_{{\rm q},1}+\tilde{\omega}_{{\rm q},2})$. Red and green 
crosses mark the perfect absorption points ($\abs{S_{11}}=0$), each 
corresponding to a one-dimensional Hermitian subspace of $H_{\rm PA}$. The 
vertical purple dashed line ($\tilde{\Delta}_s$) identifies the optimal detuning 
maximizing the efficiency for the upper hybrid mode. 
(b) Individual conversion efficiencies $\eta_1$ and $\eta_2$ as functions of $\omega_d$ at 
fixed $\Delta_\omega$ (see legend). 
(c) Emission spectrum $\tilde{S}_{{\rm q},1}(\omega)$ [see \eqref{emiss.spec}], 
normalized to its global maximum, evaluated at the detuning and drive frequency maximaxing the efficiency for the upper hybrid mode [see $\tilde\Delta_s$ in (a)], with 
$\tilde{\omega}_{{\rm q},2} = 3\,\tilde{\omega}_{{\rm q},1}$. The labeled cascade 
transitions follow the level structure of \figref{fig_avoided_crossing}(b).
All panels are calculated with 
$\gamma^{\rm r}/\tilde{\omega}_r =\gamma^{\rm qr}_{1,2}/\tilde{\omega}_r = 2.3 \times 10^{-4}$, 
$g_{\rm eff}/\tilde{\omega}_r = 1 \times10^{-3} $, 
$\gamma^{\rm nr}/\tilde{\omega}_r = 1 \times 10^{-5}$, 
$\gamma^{\rm qnr}_1/\tilde{\omega}_r = 2 \times10^{-5} $, 
$\gamma^{\rm qnr}_2/\tilde{\omega}_r = 1.25 \times10^{-5} $, 
and $|A_d| = 0.01\times\,\gamma^{\rm r}$.}
    \label{fig13}
\end{figure}

Figure~\ref{fig13}(a) shows the reflection spectra as a function of the 
detuning $\Delta_{\omega}=\tilde\omega_{r}-(\tilde{\omega}_{{\rm q},1}
+\tilde{\omega}_{{\rm q},2})$. We consider a configuration including all 
decay rates, radiative and non-radiative, for both the resonator and qubits, with the loss-balance condition deliberately broken: 
$ \gamma^{\rm r}\neq \gamma^{\rm nr}+\gamma^{\rm qr} + \gamma^{\rm qnr}$ (see \tabref{table1} for the 
definitions of the relevant physical quantities). We work in a regime where the tunable radiative losses dominate 
over the non-radiative ones. The detuning $\Delta_{\omega}$ is scanned by 
varying both dressed qubit frequencies while maintaining 
$\tilde{\omega}_{{\rm q},2}=3\,\tilde{\omega}_{{\rm q},1}$. All remaining 
parameters are listed in the figure caption.
The reflection map exhibits two PA points (red and green crosses) 
at opposite values of $\Delta_\omega$. Each PA point coincides with the 
emergence of a Hermitian subspace of the effective 
non-Hermitian Hamiltonian $H_{\mathrm{PA}}$ [see \eqref{Hrz}]: at the 
corresponding detunings, the effective loss rate of one hybrid mode vanishes, 
$\tilde{\gamma}_j=0$, owing to its dependence on the Hopfield coefficients 
[see \eqref{gammabar}]. As a result, the associated $H_{\mathrm{PA}}$ eigenvalue, $\tilde\Omega_j$, becomes purely real and a reflection zero appears when the drive is resonant with that eigenfrequency, $\omega_d = \tilde\Omega_j$. Thus, the theoretical framework of \secref{onephtwatoms} and \secref{theory_s} 
extends to this nondegenerate case upon the replacement 
$2\,\tilde\omega_q \to \tilde{\omega}_{{\rm q},1}+\tilde{\omega}_{{\rm q},2}$.

Figure~\ref{fig13}(b) shows the individual conversion efficiencies 
$\eta_j(\omega_d, \Delta_\omega)$ [see \eqref{eff}] for both qubits as the drive is scanned 
across the hybrid-mode resonances, at the optimal detuning $\tilde{\Delta}_s$ indicated by the 
vertical dashed line in panel (a). 
The global maxima of $\eta_j$ do not fall exactly on the $\abs{S_{11}}=0$ points: the competition between efficient power transfer (PA) and minimized effective non-radiative losses is detuning-dependent. Hence, the conversion efficiency peaks and the reflection zeros are slightly offset in the $(\omega_d, \Delta_\omega)$ space, as shown by the vertical purple dashed line ($\tilde\Delta_s$) and the red cross in \figref{fig13}(a). This behavior is consistent with the results in \secref{resPT} 
and \secref{Hrzsubsp}, and with the formal analysis in \appref{app_eff_PA}. The peak conversion efficiencies reach approximately $92\%$ for qubit~2 and $90\%$ for qubit~1. This small asymmetry follows directly from the choice $\gamma^{\rm qnr}_{1}>\gamma^{\rm qnr}_{2}$.

Figure~\ref{fig13}(c) displays the emission spectrum $\tilde{S}_{{\rm q},1}$, defined in \eqref{emiss.spec} and evaluated at the detuning and drive frequency corresponding to the maximum conversion efficiency for the upper hybrid mode 
(see $\tilde\Delta_s$ in \figref{fig13}(a)), with $\tilde{\omega}_{{\rm q},2} = 3\,\tilde{\omega}_{{\rm q},1}$. As in the case 
of equal qubit frequencies, the spectrum shows two peaks associated with the 
spontaneous down-conversion cascade, labeled according to the level structure 
in \figref{fig_avoided_crossing}(b) and corresponding to the transitions marked by the downward orange ($\omega_{+,1_{+}}$) and purple ($\omega_{1_-,0}$) arrows in that figure. The emission spectrum of 
qubit~2 would instead exhibit the transitions indicated by the blue ($\omega_{+,1_{-}}$) and purple ($\omega_{1_{+},0}$) 
arrows in the same figure. Unlike the degenerate case, the two qubits emit at different 
frequencies, reflecting the richer level structure of Fig.~2(b). 
This is the main qualitative difference arising when the 
frequency-degeneracy condition $\tilde{\omega}_{q,1} = \tilde{\omega}_{q,2}$ 
is lifted.

Overall, the nondegenerate configuration of qubits preserves the central conclusion of \secref{Hrzsubsp}: PA remains the spectral condition that optimizes the conversion of a single photon into two-qubit excitations, while detuning provides a robust and experimentally accessible means to engineer Hermitian subspaces of $H_{\mathrm{PA}}$ and mitigate the impact of unbalanced non-radiative losses.

\newpage
\bibliography{biblio_def}
\end{document}